\documentclass[a4paper,11pt]{article} 
\usepackage{jheppub}
\usepackage{amsmath,amssymb}
\usepackage{booktabs}
\usepackage[utf8]{inputenc}
\usepackage{hhline}
\usepackage{diagbox}
\usepackage{multirow}
\usepackage{subfigure}
\usepackage{float}
\usepackage{tabularx}
\usepackage[dvipsnames]{xcolor}
\usepackage[normalem]{ulem}
\usepackage{cancel}
\newcommand{\ba}{\begin{eqnarray}}
\newcommand{\ea}{\end{eqnarray}}
\newcommand{\be}{\begin{equation}}
\newcommand{\ee}{\end{equation}}
\newcommand{\bal}{\begin{align}}
\newcommand{\eal}{\end{align}}

\newcommand{\dd}{\mathrm{d}}
\newcommand{\barr}{\begin{eqnarray}}
\makeatletter
\def\fmslash{\@ifnextchar[{\fmsl@sh}{\fmsl@sh[0mu]}}
\def\fmsl@sh[#1]#2{%
  \mathchoice
    {\@fmsl@sh\displaystyle{#1}{#2}}%
    {\@fmsl@sh\textstyle{#1}{#2}}%
    {\@fmsl@sh\scriptstyle{#1}{#2}}%
    {\@fmsl@sh\scriptscriptstyle{#1}{#2}}}
\def\@fmsl@sh#1#2#3{\m@th\ooalign{$\hfil#1\mkern#2/\hfil$\crcr$#1#3$}}
\makeatother

\newcommand{\GeV}{\mbox{GeV}}
\newcommand{\MeV}{\mbox{MeV}}

\newcommand{\mbMSbar}{\overline{m}_b(\overline{m}_b)}

\newcommand{\refapp}[1]{appendix~\ref{app:#1}}
\newcommand{\refeq}[1]{eq.~(\ref{eq:#1})}
\newcommand{\reffig}[1]{figure~\ref{fig:#1}}
\newcommand{\reftab}[1]{table~\ref{tab:#1}}
\newcommand{\refsec}[1]{section~\ref{sec:#1}}

\definecolor{darkgreen}{RGB}{90,150,50}
\definecolor{brown}{RGB}{150,50,0}

\title{\boldmath 
The $\bar{B}\to \pi$ form factors from QCD and their impact on $|V_{ub}|$
}

\author[a]{Domagoj Leljak,}
\author[a]{Bla\v zenka  Meli\'c,}
\author[b]{Danny van Dyk}
\affiliation[a]{Rudjer Boskovic Institute, Division of Theoretical Physics, Bijeni\v cka 54, HR-10000 Zagreb, Croatia}
\affiliation[b]{Technische  Universit\"at  M\"unchen,  James-Franck-Straße  1,  85748  Garching, Germany}

\emailAdd{%
domagoj.leljak@irb.hr,
melic@irb.hr,
danny.van.dyk@gmail.com%
}

\preprint{
\begingroup
\raggedleft
EOS-2021-02\\
TUM-HEP-1316/21\\
RBI-ThPhys-2021-1\\
\endgroup
}

\abstract{
We revisit light-cone sum rules with pion distribution amplitudes to determine the full set of local $\bar{B} \to \pi$
form factors. To this end, we determine all duality threshold parameters from a Bayesian fit for the first time.
Our results, obtained at small momentum transfer $q^2$, are extrapolated to large $q^2$ where they
agree with precise lattice QCD results. We find that a modification to the commonly used BCL parametrization is
crucial to interpolate the scalar form factor between the two $q^2$ regions.
We provide numerical results for the form factor parameters -- including their covariance --- based
on simultaneous fit of all three form factors to both the sum rule and lattice QCD results.
Our predictions for the form factors agree well with measurements of the $q^2$ spectrum of the
semileptonic decay $\bar{B}^0\to \pi^+\ell^-\bar\nu_\ell$. From the world average of the latter we obtain
$|V_{ub}| = (3.77 \pm 0.15)\cdot 10^{-3}$, which is in agreement with the most recent inclusive determination at the $1\,\sigma$ level.
}
\keywords{B-decays, QCD, sum rules, CKM matrix elements}

\allowdisplaybreaks

\begin{document}

\maketitle

\flushbottom

\section{Introduction}

The world average of the inclusive $|V_{ub}|$ determinations following the BNLP approach~\cite{%
Bosch:2004th,%
Bosch:2004cb,%
Lange:2005yw,%
Neubert:2004sp,%
Neubert:2005nt%
} and the GGOU approach~\cite{%
Gambino:2007rp,%
Gambino:2011cq%
} as determined by the HFLAV collaboration reads~\cite{Amhis:2019ckw}:
\begin{equation}
\begin{aligned}
    10^3 \times |V_{ub}|_\text{BLNP}
        & = 4.44 \, {}^{+0.13}_{-0.14} |_\text{exp.} \, {}^{+0.21}_{-0.22} |_\text{theory}
          \simeq 4.44 {}^{+0.25}_{-0.26}\,,\\
    10^3 \times |V_{ub}|_\text{GGOU}
        & = 4.32 \, \pm 0.12|_\text{exp.} \, {}^{+0.12}_{-0.13} |_\text{theory}
          \simeq 4.32 {}^{+0.17}_{-0.18}\,.
\end{aligned}
\end{equation}
These results deviate significantly from $|V_{ub}|$ determinations that use the exclusive decays
$\bar{B}^0\to \pi^+\ell^-\bar\nu_\ell$, where $\ell=e,\mu$. The present world average thereof reads~\cite{Amhis:2019ckw}:
\begin{equation}
\begin{aligned}
    10^3 \times |V_{ub}|^\text{$\bar{B}\to \pi$}_\text{LQCD+LCSR}
        & = 3.67 \pm 0.09 |_\text{exp.} \, \pm 0.12 |_\text{theory}
          \simeq 3.67 \pm 0.15\,.
\end{aligned}
\end{equation}
Assuming the inclusive and exclusive results to be uncorrelated and normally distributed with the stated
overall uncertainties, these results are mutually incompatible. One finds a deviation of $\approx 2.7\sigma$, depending on which of the inclusive determinations is considered.
This long-standing situation is commonly referred to as the ``exclusive vs inclusive'' puzzle, which continues
to be a topic of active research~\cite{Gambino:2020jvv}.\\

The most recent inclusive determination by the Belle collaboration \cite{Cao:2021xqf} finds the tension reduced,
with the central value dropping closer to the exclusive one, while simultaneously increasing the uncertainty.
The average of the values extracted using four different theoretical frameworks is reported as:
\begin{equation}
    \label{eq:Vub-Belle-inclusive-2020}
    10^3 \times |V_{ub}| = 4.10 \pm 0.09 \pm 0.22 \pm 0.15\,.
\end{equation}
where the uncertainties are of statistical, systematical, and theoretical origin, respectively.
Compared to the relative uncertainty of $\approx 4\%$ in the determination from exclusive
$\bar{B}\to \pi\ell^-\bar\nu_\ell$ decays, the inclusive determination has a much larger relative
uncertainty of $\approx 7\%$. The latter is partially dominated by the subtraction of a large
$B\to X_c\ell^-\bar\nu_\ell$ background, which is one focus of the recent Belle analysis~\cite{Cao:2021xqf}.
The smallness of the (theory) uncertainties in the exclusive determination therefore warrant heightened scrutiny.\\

The description of exclusive semileptonic decays requires knowledge of the hadronic form factors.
The set of form factors includes $f_+$ and $f_0$, which are relevant to the SM predictions
for charged-current semileptonic $\bar{B}\to \pi \ell^-\bar\nu_\ell$ decays.
Another form factor $f_T$ is needed for SM predictions of rare semileptonic $\bar{B}\to \pi\ell^+\ell^-$ decays
and also arises in Beyond the Standard Model (BSM) analyses of the
charged-current decay. 
All three form factors are scalar-valued coefficients that emerge in the Lorentz decomposition
of the two hadronic matrix elements
\begin{equation}
    \label{eq:B-to-pi-FF-basis}
    \begin{aligned}
        \langle  \pi(k) | \bar{u} \gamma^\mu b | {B(p)} \rangle
            & = f_+(q^2)\, \left[(p + k)^\mu - \frac{m_B^2 - m_\pi^2}{q^2} q^\mu\right]
              + f_0(q^2)\, \frac{m_B^2 - m_\pi^2}{q^2} q^\mu\,, \\
        \langle  \pi(k)  | \bar{u} \sigma_{\mu\nu} q^{\nu} b | {B(p)} \rangle 
            & = \frac{i\, f_T(q^2)}{m_B + m_\pi} \left[q^2 (p + k)_{\mu} - \left (m_B^2 - m_\pi^2 \right ) q_{\mu}    \right ] \, .
    \end{aligned}
\end{equation}
These three form factors are all functions of the momentum transfer $q^2 \equiv (p - k)^2$.

Presently, the determination of $|V_{ub}|$ from the exclusive $\bar{B}\to \pi\ell^-\bar\nu_\ell$
decays is the most competetive. Other determinations either lack precision on the theoretical
side (such as $\bar{B}_c\to D\ell^-\bar\nu_\ell$) or the experimental side (such as $\bar{B}\to \ell^-\bar\nu_\ell$
or $\Lambda_b\to p\mu^-\bar\nu_\mu$), with improvements to the precision expected in the future.
A more detailed discussion is available in ref.~\cite{Amhis:2019ckw}.
The increase in precision of the theoretical predictions for and the experimental measurements of
$\bar{B}\to \pi\ell^-\bar\nu_\ell$
has also made this decay a prime candidate for searches of BSM effects in charged
currents. These searches are well motivated in light of recent tensions in $b\to c\ell^-\bar\nu_\ell$
processes.

The purpose of this work is three-fold:
\begin{enumerate}
    \item to revisit light-cone sum rule predictions for the full set of local $\bar{B}\to \pi$
form factors, with focus on the systematic uncertainties that affect this method;
    \item to carry out a combined fit with the precise lattice QCD (LQCD) results for the form factors,
in order to provide the most up-to-date exclusive determination of $|V_{ub}|$;
    \item to provide up-to-date predictions for $\bar{B}\to \pi \ell^-\bar\nu_\ell$ observables that probe
        lepton-flavour universality and non-standard weak effective couplings.
\end{enumerate}

%
%
\section{The $\bar{B}\to \pi$ form factors from light-cone sum rules}
\label{sec:lcsr}

Hadronic transition form factors, such as in $\bar{B}\to \pi$ transitions, are genuinely non-perturbative
objects. They cannot be computed with perturbative methods in the phase space region in which they are needed
to describe the semileptonic decays. 
Light-Cone Sum Rules (LCSRs) are a long-established technique to determine hadronic form factors \cite{Braun:1988qv, Balitsky:1989ry, Chernyak:1990ag}.
Within a LCSR, the hadronic transition form factor of interest is determined from a calculation
of a suitable correlation function. One can find a kinematic regime in which this correlation
function factorizes into a pertubative (hard) scattering kernel and universal nonperturbative matrix elements,
the so-called light-cone distribution amplitudes (LCDAs). Using dispersion relations and the assumption
of semi-global quark hadron duality, the sum rule then gives the form factor of interest.
A pedagogical introduction of LCSRs in particular and a modern perspective on QCD sum rules in general
can be found in ref.~\cite{Colangelo:2000dp}.
In this work we revisit the LCSRs for full set of local $\bar{B}\to \pi$ form factors associated with dimension-three
$b\to u$ or $b\to d$ currents. The LCSRs are constructed with an on-shell pion and an interpolated $B$ meson,
and by the use of pion distribution amplitudes.  Our definition of the form factors is shown in \refeq{B-to-pi-FF-basis}.\\

The analytical expressions for the two-point correlation functions that give rise to the sum rules are
known to high accuracy. The expansion in light-cone operators uses the twist of an operator --- the
difference between mass dimension and canonical spin of the operator  --- as an expansion parameter. Operators of higher
twist are supressed by power of $\Lambda_\text{had} / m_b$.
The leading contributions at the twist-two level are known at next-to-leading order (NLO) in $\alpha_s$~\cite{Ball:2004ye, Duplancic:2008ix}.
Next-to-next-to-leading order (NNLO)~\cite{Bharucha:2012wy} are partially computed in the large $\beta_0$ approximation.
In $\bar{B}\to \pi$ transitions, the next-to-leading twist contributions are known to by enhanced by the factor
\begin{equation}
    \frac{\mu_\pi}{m_b} = \frac{m_\pi^2}{m_b (m_u + m_d)}\,,
\end{equation}
which is formally power-suppressed but numerically large. Due to this enhancement, the twist-three terms
contribute approximately $50\%$ to the correlation function, e.g. \cite{Duplancic:2008ix}.
Due to the chiral enhancement it is important to include the twist-three terms also at NLO~\cite{Duplancic:2008ix}.
Beyond this level, contributions up to twist-six follow the expected pattern of power suppression~\cite{Rusov:2017chr}.\\

In this work, we provide predictions for the three hadronic form factors based on the analytic expressions
in ref.~\cite{Duplancic:2008ix, Duplancic:2008tk, Khodjamirian:2009ys}. These LCSRs use $\pi$-meson distribution amplitudes. They include expressions up to twist-four accuracy at leading order in $\alpha_s$ and 
expressions up to twist-three accuracy at next-to-leading order in $\alpha_s$. Expressions beyond twist-four
accuracy are numerically negligible~\cite{Rusov:2017chr}.
In the preparation of this work we have identified two typos in the analytic expressions in the literature.%
\footnote{
First, in eq.~(4.12) of ref.~\cite{Duplancic:2008ix} the factor $1/2$ in front of the $\dd^2 \phi_{4\pi}/\dd u^2$ term
should be replaced by a factor $1/4$. Second, in the fourth line of eq.~(B.35) the plus prescription should
extend to the entire term rather than only to the  $\rho/(1 - \rho)$ factor. The first typo is corrected in subsequent
publications, while the second typo is not.
}
These two typos do not significantly impact the form factor values, but have a non-negligible effect on the computation
of the $B$-meson mass predictor, which we use below to determine the duality thresholds.\\

Our numerical results for the form factors as presented below differ from previous LCSR studies in the following
aspects:
\begin{enumerate}
    \item We use updated input parameters for quark masses, strong coupling and --- most importantly --- for
    the two-particle twist-two $\pi$ LCDA. The full set of input parameters is discussed in \refsec{inputs}.
    
    \item We determine the duality thresholds for \emph{all three} form factors from three daughter sum rules.
    The latter are obtained from the derivative of the initial sum rules with respects to the Borel parameter.
    In this way a predictor for the mass squared of the $B$ meson can be included in a statistical analysis.
    The method is discussed for the $f_+$ form factor in ref.~\cite{Imsong:2014oqa}, and for LCSRs with
    $B$-meson LCDAs in ref.~\cite{Gubernari:2018wyi}.
    Details of this procedure and practical considerations for this step are discussed in \refsec{thresholds}.

    \item Within the threshold-setting procedure, we investigate the dependence of the duality thresholds on the
    momentum transfer $q^2$. We compare two models of these thresholds, and use their difference to assign a systematic
    uncertainty to our final results.
\end{enumerate}

\subsection{Input parameters}
\label{sec:inputs}

\begin{table}[p]
\centering
\renewcommand{\arraystretch}{1.15}
\begin{tabular}{ c  c  c  c  c }
    \toprule
    Parameter                                   & value/interval                 & unit     & prior               & comments/source \\
    \hline
    \multicolumn{5}{c}{quark-gluon coupling and quark masses}\\
    \hline
    $\alpha_s(m_Z)$                             &  0.1179  $\pm$ 0.0010          & ---      & gaussian & \cite{Zyla:2020zbs}                          \\
    $\mbMSbar$                                  &    4.18  $\pm$ 0.03            & \GeV     & gaussian & \cite{Zyla:2020zbs}                          \\
    $[m_u+m_d](2\,\GeV)$                         & 6.9 $\pm$ 1.1                  & \MeV     & gaussian & \cite{Zyla:2020zbs}                          \\
    \hline
    \multicolumn{5}{c}{hadron masses}\\
    \hline
    $m_B$                                       &  5279.58                       & \MeV     & ---      & \cite{Zyla:2020zbs}                          \\
    $m_\pi$                                     & 139.57                         & \MeV     & ---      & \cite{Zyla:2020zbs}                          \\
    \hline
    \multicolumn{5}{c}{vacuum condensate densities}\\
    \hline
    $\langle\bar{q}q (2\GeV)\rangle$            & $-(288^{+17}_{-14})^3$         & $\MeV^3$ & ---      & $m_\pi^2f_\pi^2/2(m_u+m_d)$                  \\
    $\langle\frac{\alpha_s}{\pi} G^2\rangle$    & $[0.000, 0.018]$               & $\GeV^4$ & uniform  & \cite{Ioffe:2002ee}                          \\
    $m_0^2$                                     & $[0.6, 1.0]$                   & $\GeV^2$ & uniform  & \cite{Ioffe:2002ee}                          \\
    $r_{vac}$                                   & $[0.1, 1.0]$                   & ---      & uniform  & \cite{Ioffe:2002ee}                          \\
    \hline
    \multicolumn{5}{c}{parameters of the pion DAs}\\
    \hline
    $f_\pi$                                     & $130.2$ $\pm$ $0.8$            & \MeV     & gaussian & \cite{Aoki:2019cca}                          \\
    $a_{2\pi}(1 \GeV)$                          & $0.157$ $\pm$ $0.027$          & ---      & gaussian & \cite{Bali:2019dqc}                         \\
    $a_{4\pi}(1 \GeV)$                          & $[-0.04, 0.16]$                & ---      & uniform  & \cite{Khodjamirian:2011ub}                   \\
   $ \mu_\pi(2 \GeV) $                         &  $2.8^{+0.6}_{-0.4} $           & \GeV     & ---      & $m_\pi^2/(m_u+m_d)$                          \\
    $f_{3\pi}(1 \GeV)$                          & $[0.003, 0.006]$               & $\GeV^2$ & uniform  & \cite{Ball:2006wn}                           \\
    $\omega_{3\pi}(1 \GeV)$                     & $[-2.2, -0.8]$                 & ---      & uniform  & \cite{Ball:2006wn}                           \\
    $\delta_\pi^2(1 \GeV)$                      & $[0.11, 0.33]$                 & $\GeV^2$ & uniform  & ($50\%$ sys. unc.) \cite{Bali:2019dqc}                            \\
    $\omega_{4\pi}(1 \GeV)$                     & $[0.1, 0.3]$                   & ---      & uniform  & \cite{Ball:2006wn}                           \\
    \hline
    \multicolumn{5}{c}{sum rule parameters and scales}\\
    \hline
    $\mu$                                       & $3.0$                          & $\GeV$   & ---      & \cite{Khodjamirian:2011ub, Gelhausen:2013wia} \\
    $M^2$                                       & $[12.0, 20.0]$                 & $\GeV^2$ & uniform  & \cite{Khodjamirian:2011ub}                   \\
    $s_0^{f_+}$                                 & $[30.0, 42.0]$                 & $\GeV^2$ & uniform  &                                              \\
    $s_0^{\prime\, f_+}$                        & $[-1.0, +1.0]$                 & ---      & uniform  &                                              \\
    $s_0^{f_0}$                                 & $[30.0, 42.0]$                 & ---      & uniform  &                                              \\
    $s_0^{\prime\, f_0}$                        & $[-1.0, +1.0]$                 & ---      & uniform  &                                              \\
    $s_0^{f_T}$                                 & $[30.0, 42.0]$                 & $\GeV^2$ & uniform  &                                              \\
    $s_0^{\prime\, f_T}$                        & $[-1.0, +1.2]$                 & ---      & uniform  &                                              \\
    $\overline{M}^2$                            &  $5.5 \pm 1.0$                 & $\GeV^2$ & gaussian & \cite{Gelhausen:2013wia}                     \\
    $\overline{s}_0^B$                          & $[29.0, 44.0]$                 & $\GeV^2$ & uniform  &                                              \\
    \bottomrule
\end{tabular}
\caption{Input parameters used in the numerical analysis of the two-point sum rules for the $f_B$ decay constant and LCSRs for $\bar{B} \to \pi$ form factors.
The full prior distribution is a product of uncorrelated individual priors, which are either
uniform or Gaussian distributed. Gaussian priors cover the intervals
at 68\% probability, and their central value corresponds to the mode.
For practical purpose, variates of the gaussian priors are only sampled inside their respective
99\% intervals. The prior intervals of the duality threshold parameters are chosen such that 
the peaking posterior distribution is fully contained.
}
\label{tab:inputs-sr}
\end{table}
\noindent
Our setup follow the usual approach to calculate both the $B$-meson decay constant $f_B$ in two-point
QCD sum rules and the $\bar{B}\to \pi$ form factors in LCSR within a simultaneous analysis~
\cite{Duplancic:2008ix,Imsong:2014oqa,Khodjamirian:2017fxg}.
The rationale for this approach is that perturbative corrections to the correlation functions
in both sum rules partially cancel.
As a consequence, our input parameters involve the full set of all, the two-point sum rule and the light-cone sum rule 
parameters. We classify these parameters as follows:
\begin{description}
    \item[strong coupling and quark masses] These parameters include the strong coupling
    at $\mu = M_Z$, the bottom quark mass in the $\overline{\text{MS}}$ scheme at the scale $m_b$,
    and the sum of the up and down quark masses in the $\overline{\text{MS}}$ scheme at the scale $2\,\GeV$.
    
    \item[hadron masses] These parameters include the masses of the initial-state $B$ meson $m_B$ and the final-state pion $m_\pi$.
    
    \item[vacuum condensate densities] These parameters include the quark condensate evaluated using the GMOR relation at $2\, \mathrm{GeV}$ and the gluon condensate, while the mixed quark-gluon condensate is implemented through $m_0^2$, its ratio with the quark condensate. Lastly, $r_{\mathrm{vac}}$ parametrizes factorization in the four-quark condensate density. These
    parameters are needed exclusively in the two-point sum rule.
    
    \item[parameters of the $\pi$ LCDAs] These parameters include the pion decay constant $f_\pi$ to which the leading-twist
    LCDA is normalised. The shape of the leading-twist DA is described by an expansion in Gegenbauer polynomials,
    which are eigenfunction of the RGE kernel to leading-logarithmic accuracy. Isospin symmetry implies that only
    even Gegenbauer polynomials contribute, and we retain the first two non-vanishing Gegenbauer coefficients $a_{2\pi}$ and $a_{4\pi}$.
    Following ref.~\cite{Ball:2006wn}, we normalise the twist-3 two-particle LCDAs to the chiral parameter $\mu_\pi(2\,\GeV)$ and
    twist-three three-particle LCDAs to the decays constants $f_{3\pi}$. The shape of the three-particle LCDAs additional involves
    the parameter $\omega_{3\pi}$. The twist-four LCDAs  are parametrized in terms of $\delta^2_\pi$ and $\omega_{4\pi}$. If not
    specified otherwise, all parameter of this class are renormalised at a scale of $1\,\mathrm{GeV}$.
    
    \item[sum rule parameters and scales] These parameters include the Borel parameter $M^2$ and the values and slopes
    of the duality threshold parameters $s_0^F$ and $s_0^{\prime F}$, where $F$ denotes one of the form factors $\{f_+,f_0,f_T\}$.
    We discuss the parametrisation of the thresholds in detail below. The perturbative LCSR kernels are evaluated at a renormalisation scale $\mu$.
    Further parameters are the Borel parameter $\bar{M}^2$ and duality threshold $\bar{s}_0^B$ of the
    auxilliary two-point sum rule. 
\end{description}
All the input parameters are listed and their prior probability density functions (PDFs) are summarized in \reftab{inputs-sr}.

We briefly discuss the differences between the inputs used in this work and the inputs used in refs.~\cite{Imsong:2014oqa, Khodjamirian:2017fxg}:
\begin{enumerate}
    \item While the input parameters for the light quark masses $m_u$ and $m_d$ change only slightly,
        this change has a large numerical effect on $\mu_\pi^2$, which normalises the twist-three two-particle
        contributions to the sum rules. It also affected the value of $\bar{q}q$ condensate density.

    \item A recent lattice QCD analysis~\cite{Bali:2019dqc} of the shape of the leading-twist pion LCDA has
        provided for the first time a determination of the leading Gegenbauer moment $a_2$ from first principles.
        We use this result as a Gaussian prior in our analysis. Note that we use the RGE to LL to translate the
        lattice results to our default input scale of $1\,\mathrm{GeV}$.
        We also adjust the uniform prior PDFs for the parameters $a_4$ and $\delta_\pi^2$ to match the lattice
        QCD results for these parameters within their uncertainty intervals.
        
    \item We slightly increase the Borel window for the LCSR to the interval $12\,\GeV^2 \leq M^2 \leq 20\,\GeV^2$
        in which we vary the Borel parameter uniformly rather than with a gaussian prior. This increases the uncertainty
        due to the Borel parameter in the final numerical results and also fully includes the peaking structure in the 
        posterior PDF.
\end{enumerate}

\subsection{Setting the duality thresholds and Borel parameters}
\label{sec:thresholds}

Each of the duality thresholds $s_0^F$ corresponds to a point at which to artificially split the dispersive integral for its form factor $F$ into two contributions: one corresponding to the $\bar{B}\to \pi$ form factor, and
one corresponding to hadronic transition matrix elements for excited $B$-mesons and the continuum of $b$-flavoured
states. 
To obtain the threshold parameters, one commonly uses daughter sum rules obtained by taking a derivative
of the form factors' correlation function with respect to $-1/M^2$ and by subsequently normalizing
to the correlation function. By using the same input parameters as in the original sum rule, one thereby constrains
the duality thresholds parameters. This new daughter sum rule can be cast into a pseudo observable 
that serves as a predictor of the mass square for the interpolated state, i.e. here the $B$ meson;
see e.g. ref.~\cite{Imsong:2014oqa}. Schematically, 
\begin{equation}
    \label{eq:MB2-LCSR}
    [m_B^2(q^2; F)]_\text{LCSR} 
        = \frac{\int_0^{s_0} ds \,s \,\rho^F(s, q^2) \,e^{-s/M^2}}{\int_0^{s_0} ds \,\rho^F(s, q^2)\, e^{-s/M^2}}\,.
\end{equation}
Here $F$ denotes any of the three form factor under consideration, and $\rho^F$ is the OPE result for the form factor's spectral density.

To determine the duality thresholds we follow the procedure used in ref.~\cite{Imsong:2014oqa} for the $f_+$ form factor.
We construct a theoretical Gaussian likelihood centered around the experimental results for the $B$-meson mass. We further assign
a theoretical uncertainty of $1\%$ to the LCSR prediction of the $B$-meson mass. For each form factor the likelihood
challenges the LCSR predictions for the mass in five different $q^2$ points equally spaced between $-8\,\GeV^2$
to $+8\,\GeV^2$. We then fit the parameters listed in \reftab{inputs-sr} to this likelihood, using
two different models for the duality thresholds, see below. The posteriors for
most parameters are in good agreement with the priors, with the exception of the posteriors for the duality threshold parameters
and the LCSR Borel parameter, which change from uniform to peaking distributions. This change clearly
indicates that we successfully infer the duality thresholds and the Borel parameter from the
daughter sum rules.
\begin{figure}[t]
    \centering
    \begin{tabular}{cc}
        \includegraphics[width=.49\textwidth]{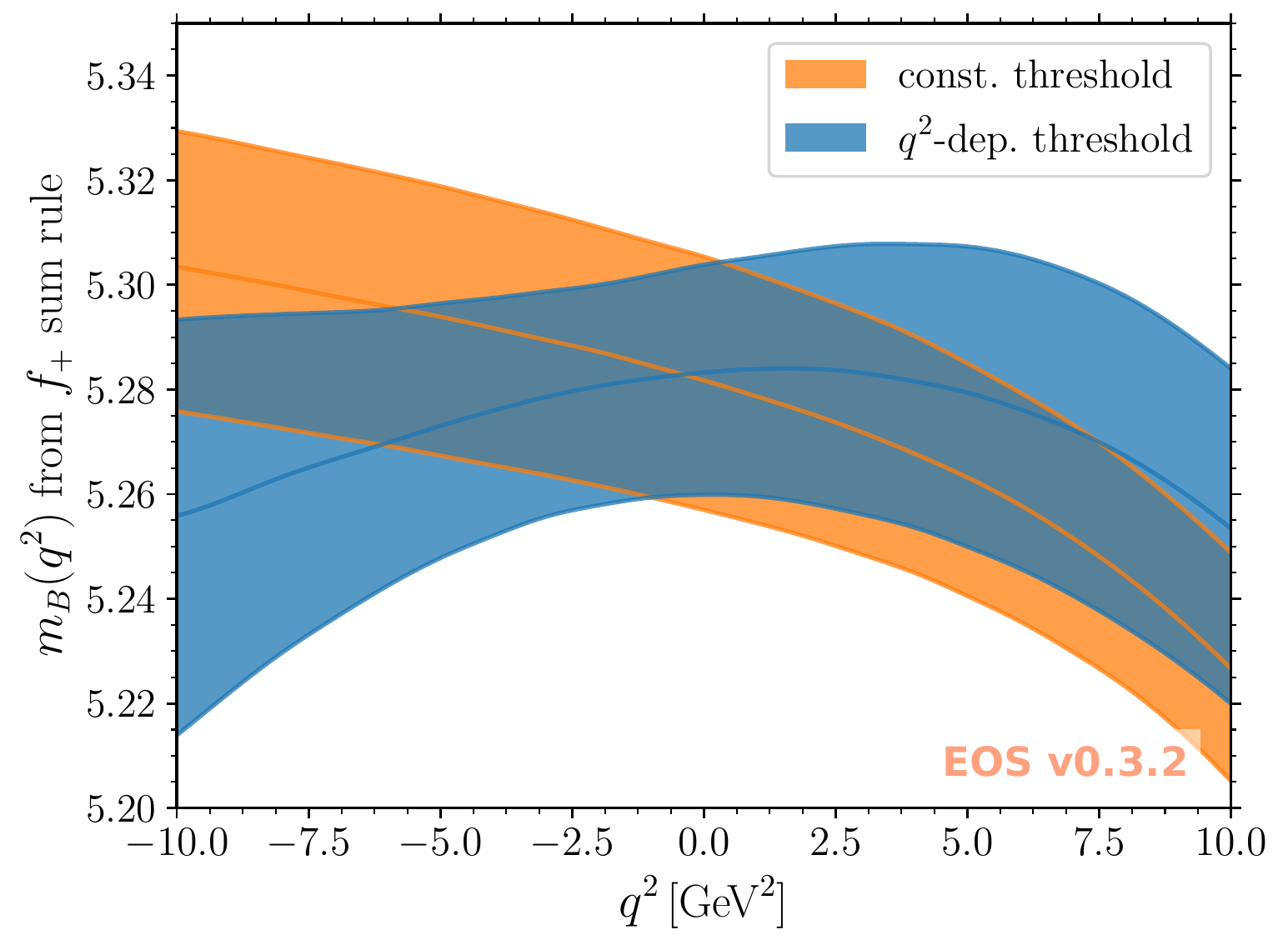} &
        \includegraphics[width=.49\textwidth]{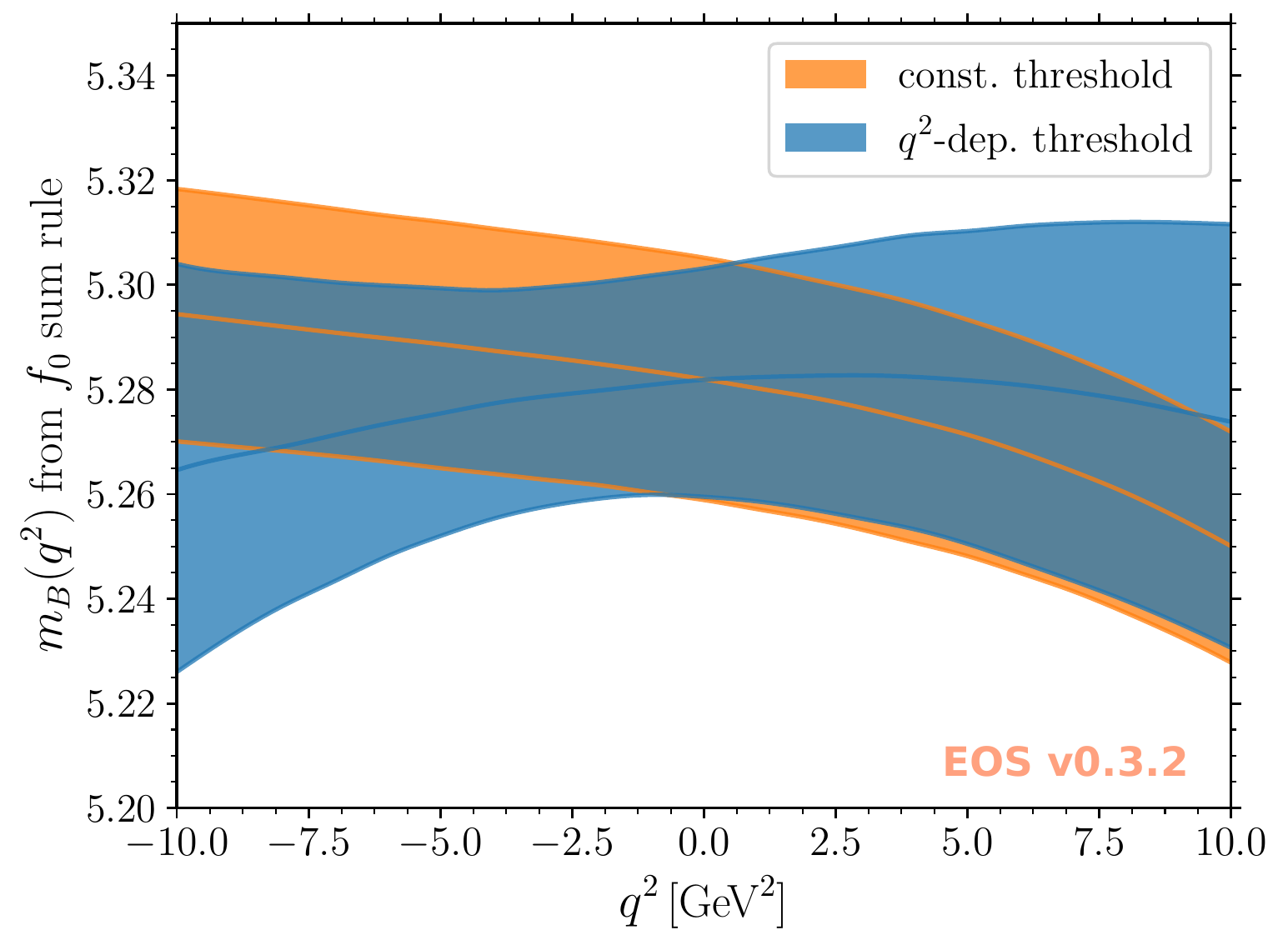} \\
        \includegraphics[width=.49\textwidth]{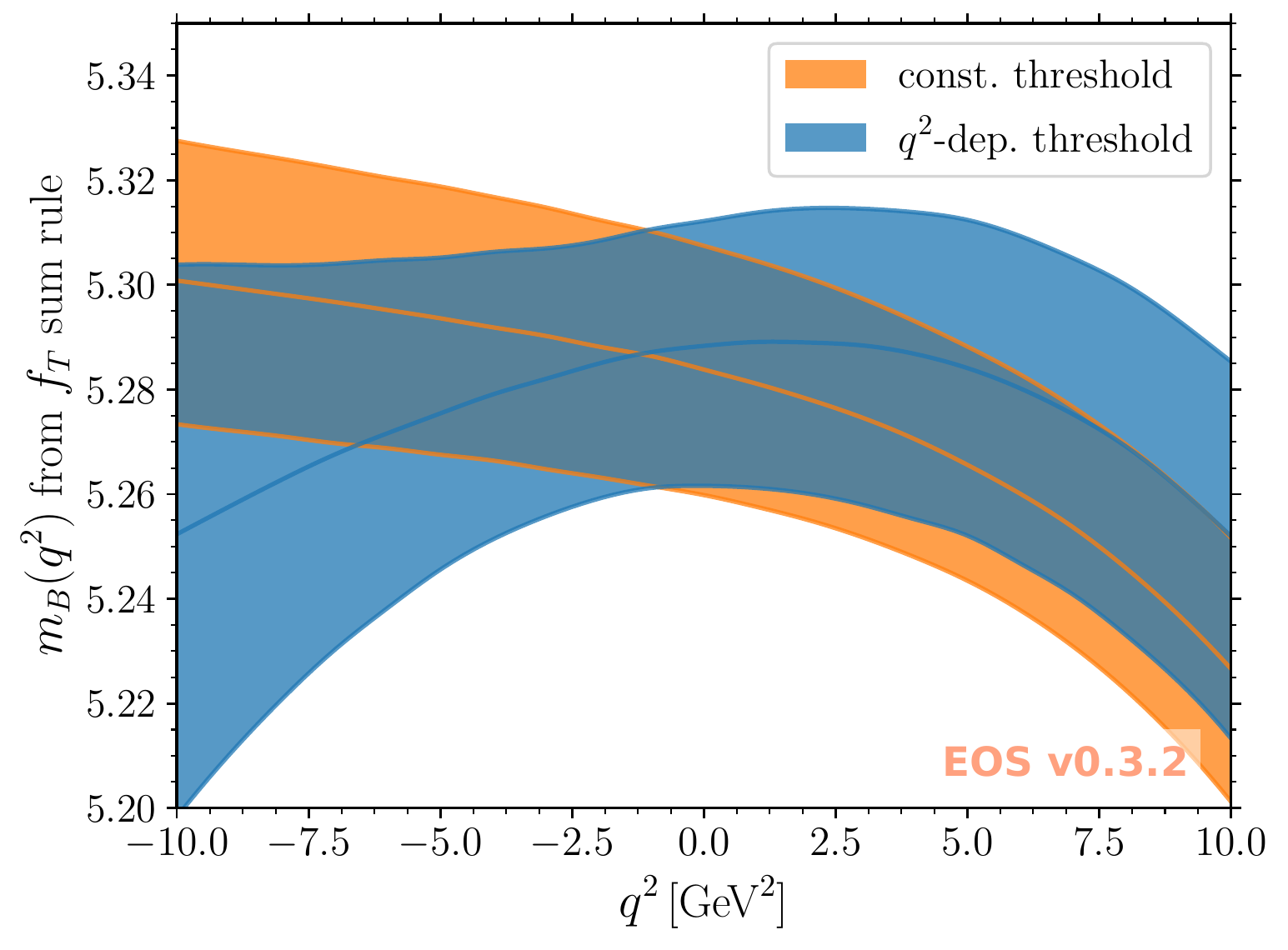} &
        \raisebox{\bigskipamount}{
        \begin{minipage}[b]{.45\textwidth}
            \caption{
                The dependence of the $B$-meson mass predictor
                $[m_B^2(q^2; F)]_\text{LCSR}$
                for each of the three form factors $F = \{f_{+}, f_{0}, f_T\}$
                on the momentum transfer $q^2$. We show the posterior-prediction for a $q^2$-invariant
                threshold (in orange) and a threshold with linear $q^2$ dependence as in eq.~\eqref{eq:s0-ansatz} (in blue).
                The shaded areas correspond to the respective $68\%$ probability envelopes.
            }
            \label{fig:q2-dep-fp}    
        \end{minipage}
        }
    \end{tabular} 
\end{figure}

The procedure carried out in this work is similar but not identical to the one presented in~\cite{Imsong:2014oqa}.
It differs in the following points:
\begin{enumerate}
    \item We determine all three transition form factors simultaneously, while in ref.~\cite{Imsong:2014oqa} the analysis is constrained to $f_+$ only. Our procedure restricts the possible parameters space more strongly, since
    all form factors share the same input parameter set except for their respective threshold parameters.
    The effect is mostly visible in the posterior of the Borel parameter and discussed in detail below.
    
    \item We \emph{do not} determine the $q^2$ derivatives of the form factors as suggested in ref.~\cite{Imsong:2014oqa}.
    Our decision is based on the following observation. If the predictor for a form factor and for its $q^2$ derivative
    share the same threshold parameter, then the mass predictor for the derivative cannot in general be expected to produce
    a value close to the $B$ meson mass squared. We would therefore need to introduce new and independent duality threshold
    parameters for each derivative. This reduces the usefulness of the derivatives as we can extract
    a similar amount of information by increasing the number of $q^2$ points, which is computationally easier.\\
\end{enumerate}

In a first fit we assume the duality thresholds to be constant with respect to $q^2$. In a second fit,
we allow for a linear $q^2$ dependence of the thresholds, i.e.,
\begin{equation}
    \label{eq:s0-ansatz}
    s_0^F(q^2) \equiv s_0^F + q^2\, s_0^{\prime F}\,.
\end{equation}
As already disussed in ref.~\cite{Imsong:2014oqa}, we find evidence for a
mild $q^2$ dependence of the duality thresholds. Here, we find a reduction of the
global $\chi^2$ by $\sim 0.5$ when allowing for a linear $q^2$ dependence in
all three thresholds. This has to be compared to a decrease of three degrees of freedom.
While this result does in no way \emph{require} to impose a $q^2$ dependence
of the thresholds, we consider it grounds enough to further investigate the
$q^2$ dependence of the $B$-meson mass predictors. To this end, we compute
the median curve and its $68\%$ probability envelope for each mass predictor
and for both fit models.
Our findings are illustrated in \reffig{q2-dep-fp}. 
\begin{figure}[t]
    \centering
    \includegraphics[width=0.7\textwidth]{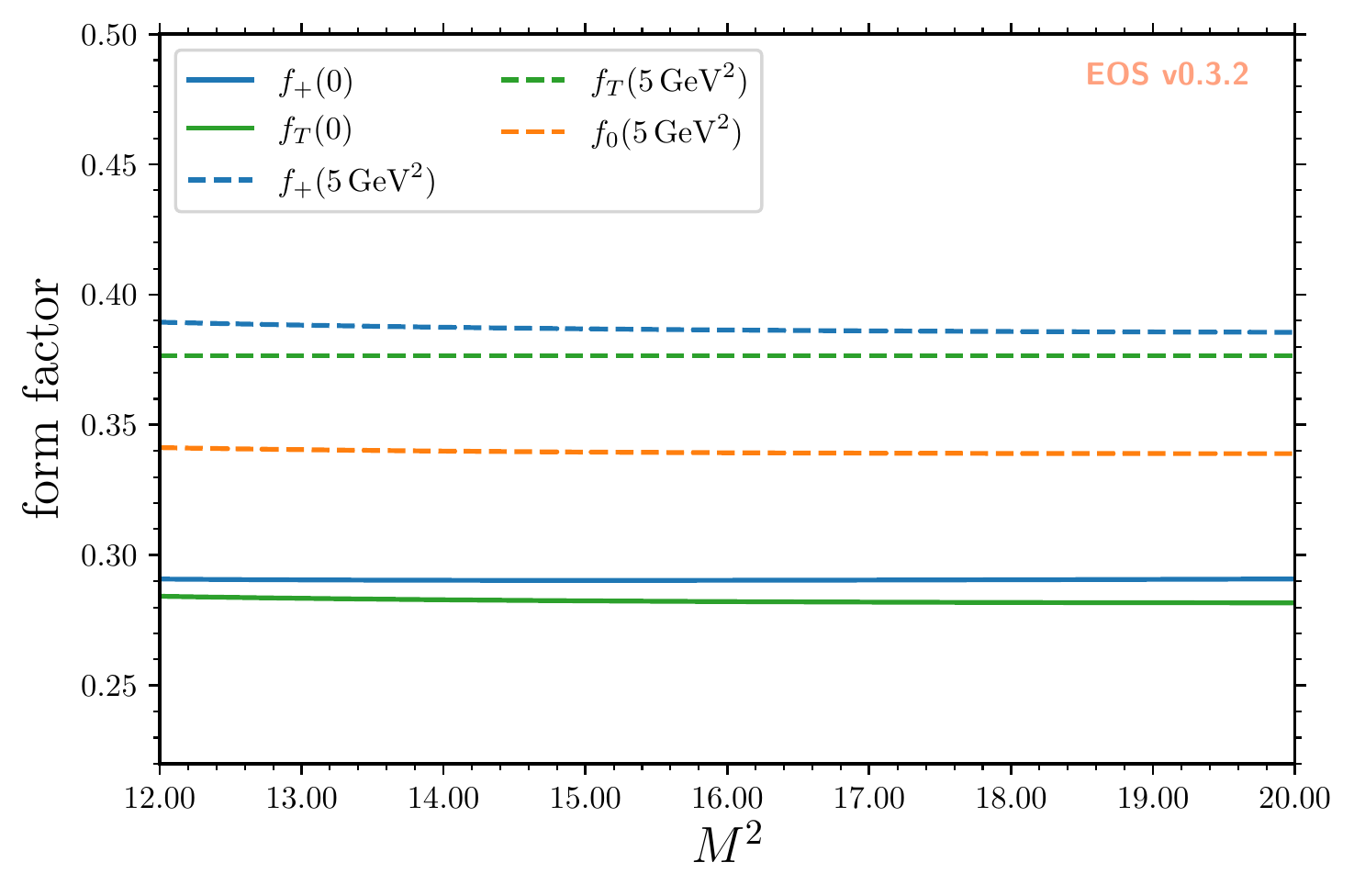}
    \caption{
        The dependence of the form factors $f_+$ (in blue), $f_0$ (in orange), and $f_T$ (in green)
        on the Borel parameter $M^2$. We exemplify this dependence for two different choices of $q^2$: $0\,\GeV^2$ (solid lines),
        and $5\,\GeV^2$ (dashed lines). The form factor $f_0$ coincides with $f_+$ at $q^2 = 0$, and is therefore
        not depicted at this value.
        }
    \label{fig:borel-dep}
\end{figure}
As can be expected due to the three
additional parameters, the $68\%$ envelopes of the fit model with $q^2$-dependent
thresholds (blue bands) have a larger uncertainty than the envelopes of the
fit model with constant thresholds (orange bands). However, we also find that the former model
reproduces the physical $B$-meson mass on average better than the fit model with constant thresholds
in the $q^2$ interval considered here. We further find that the
maximal deviation of the $B$-meson mass predictors from the physical
mass is reduced for all three form factors. We therefore chose the $q^2$-dependent
ansatz for the duality thresholds for the central values of our form factor predictions.
The difference between the constant and the $q^2$-dependent threshold parametrizations is used
to estimate systematic uncertainties due to the determination of the duality threshold parameters.

We account for the dependence on the Borel parameter $M^2$ by varying this parameter in our
prefered window $12\,\GeV^2 \leq M^2 \leq 20\,\GeV^2$. In this way,
we account for the residual $M^2$ dependence of the form factor predictions. This
procedure was carried out in ref.~\cite{Imsong:2014oqa}, where a Gaussian prior was used.
Here, we apply this procedure instead with a uniform prior.
Despite the mild dependence of each form factor on the Borel parameter value, we find that its
posterior differs strongly from its prior, with a peak at around $15\,\GeV^2$. This can be understood,
since each form factor and each $q^2$ point entering the theoretical likelihood differs slightly
in its dependence on the Borel parameter. Only when investigating all form factors and $q^2$ simultaneously,
we find that the posterior of the Borel parameter exhibits a clearly peaking structure.
The overall form factors dependence on the Borel parameter is very weak and is shown in ~\reffig{borel-dep}.

\subsection{Numerical results for the form factors}

We proceed to predict the form factors at five equally-distanced $q^2$ points
in the interval $-10\,\GeV^2 \leq q^2 \leq +10\,\GeV^2$. Our choice of points simultaneously maximizes
the number of pseudo data points while keeping correlations of neighbouring points below
$80 \%$ in the combined parametric and systematic uncertainty.
Two systematic uncertainties are estimated by the following procedures:
\begin{enumerate}
    \item For each form factor prediction the renormalization scale is varied by dividing and multiplying it with a factor of $1.25$,
    corresponding to the interval $[2.40\,\GeV, 3.75\,\GeV]$. We find that the maximal one-sided variation of all our predictions
    can be found when lowering the renormalization scale. Across all form factors and all $q^2$ points, this variation evaluates
    consistently to $\sim 4\%$. For a conservative estimate of this effects, we add an uncorrelated $4\%$ systematic uncertainty
    to all form factor predictions.
    \item For each form factor we compute the difference between the predictions with constant duality thresholds
    and $q^2$-dependent duality thresholds.
    We find that the largest difference occurs for $f_0(q^2 = 10 \, \mathrm{GeV}^2)$,
    corresponding to roughly $\sim 6\%$ of the central value. We add the differences in quadrature to the variances.
\end{enumerate}
\noindent The joint posterior predictive distribution for all of the form factors is to excellent approximation a multivariate Gaussian
distribution. We provide the mean values and standard deviations in \reftab{ff-results}.
The correlation matrix is provided in \refapp{lcsr-corr}. For convenience the mean values and the covariance matrix
are attached to the arXiv preprint of this manuscript as an ancillary machine-readable file.
One can immediately notice the very close numerical values of $f_+$ and $f_T$ form factors,
which is expected as a consequence of the heavy-quark expansion and the large-energy symmetry limit~\cite{Beneke:2000wa}.
The $f_0(q^2=0)$ pseudo data point is not included, since it coincides with $f_+(q^2=0)$ by definition.
We consistently find uncertainties $\sim 10\%$ across all $q^2$ points.
The parametric covariance matrix for our results exhibit a large degree of correlation.
The determinant of $\rho$, the linear correlation matrix reads
\begin{equation}
    \det \rho\bigg|_{\text{parametric}} = 4.0 \times 10^{-31}\,.
\end{equation}
Accounting for the systematic uncertainties as discussed above increases the determinant
to
\begin{equation}
    \det \rho\bigg|_{\text{total}} = 3.7 \times 10^{-5}\,,
\end{equation}
thereby reducing the degree of correlation. The largest correlation of $81 \%$ occurs amongst $f_0(-5\, \mathrm{GeV}^2)$ and $f_+(-5\, \mathrm{GeV}^2)$.
These findings give confidence that the 14 data points can be treated as 14 independent observations
in the following studies.

\begin{table}
    \begin{tabular}{ c r @{\,$\pm$\,} l r @{\,$\pm$\,} l r @{\,$\pm$\,} l r @{\,$\pm$\,} l r @{\,$\pm$\,} l}
    \toprule
    $q^2$
        & \multicolumn{2}{c}{$-10\,\GeV^2$}
        & \multicolumn{2}{c}{$-5\,\GeV^2$}
        & \multicolumn{2}{c}{$0\,\GeV^2$}
        & \multicolumn{2}{c}{$+5\,\GeV^2$}
        & \multicolumn{2}{c}{$+10\,\GeV^2$}
        \\
    \midrule
    $f_+(q^2)$
        & $0.170$ & $0.022$
        & $0.224$ & $0.022$
        & $0.297$ & $0.030$
        & $0.404$ & $0.044$
        & $0.574$ & $0.062$
        \\
    $f_0(q^2)$
        & $0.211$ & $0.029$
        & $0.251$ & $0.024$
        & \multicolumn{2}{c}{---}
        & $0.356$ & $0.040$ 
        & $0.441$ & $0.052$
        \\
    $f_T(q^2)$
        & $0.170$ & $0.021$
        & $0.222$ & $0.020$
        & $0.293$ & $0.028$
        & $0.396$ & $0.039$
        & $0.560$ & $0.053$
        \\
    \bottomrule
    \end{tabular}
    \caption{
        The LCSR predictions for the form factors in five $q^2$ points. The value of $f_0(q^2 = 0)$ is not independent, since
        $f_+(q^2 = 0) = f_0(q^2 = 0)$ by construction.
    }
    \label{tab:ff-results}
\end{table}

\section{Extrapolation of the LCSR results to large $q^2$}
\label{sec:extrapolation}

A central elements to the LCSR calculation of the form factors is the expansion of a suitable two-point correlation
function in terms of bilocal operators with light-like separation. This expansion is called a light-cone operator product
expanion (LCOPE). The light-cone dominance of the OPE crucially depends on the kinematic variables. It has been shown that
the light-cone dominance holds for a four-momentum transfer~\cite{Duplancic:2008ix}
\begin{equation}
    q^2 < m_b^2 - 2 m_b \bar{\Lambda} \sim 15\, \GeV^2\,.
\end{equation}
For phenomenological applications --- such as theory predictions for the total $\bar{B}\to \pi\ell^-\bar\nu$ branching fraction
and the comparison to lattice QCD results of the form factors --- we need to extrapolate our LCSR results
to $q^2$ values for which light-cone dominance does not hold. 
The standard approach to extrapolate the form factors is a fit of a parametrization of the form factors to LCSR pseudo data points.
There are several competing parametrizations, and there is no clear and an objectively preferred choice.
For the extrapolation of our LCSR results to large $q^2$ we choose the BCL parametrization~\cite{Bourrely:2008za} as it is commonly applied in the literature~\cite{Lattice:2015tia, Bailey:2015nbd, Flynn:2015mha}.\\

The BCL parametrization is based on an expansion of the form factor in the variable
\begin{equation}
    z(q^2; t_+, t_0) = \frac{\sqrt{t_+-q^2} - \sqrt{t_+-t_0}}{\sqrt{t_+-q^2} + \sqrt{t_+-t_0}}\,,
\end{equation}
where $t_+ \equiv (m_B + m_\pi)^2$ represents the $B\pi$ pair-production threshold, and $t_0 < t_+$ is a free parameter.
The semileptonic phase space $0 \leq q^2 \leq t_- \equiv  (m_B - m_\pi)^2$ is mapped onto the real $z$ axis.
The magnitude of $z(q^2)$ for $q^2$ within the semileptonic phase space is minimized by choosing
\begin{equation}
    t_0 = t_{0,\mathrm{opt}} = (m_B + m_\pi)(\sqrt{m_B}-\sqrt{m_\pi})^2\,,\\
\end{equation}
which we also adopt here. As a consequence,  $|z| < 0.284$ for semileptonic $\bar{B} \to \pi$ decays.
The $\bar{B}\to \pi$ vector and tensor form factors feature a single subthreshold pole, each due to the
$B^*$ bound state, which is located outside the semileptonic phase space. The scalar form factor $f_0$
has no subthreshold pole.

The BCL parametrization simultaneously encodes the correct asymptotic behaviour of the vector form factor $f_+$
in the limit $q^2 \to \infty$ and accounts for its subthreshold pole through a factor $(1 - q^2/M_{B^*}^2)^{-1}$.
The remainder of the form factor is then Taylor expanded in the variable $z$ to some
order $K$, with expansion coefficients $a_n$.
Accounting for the known subthreshold pole accelerates the convergence of the series.\\
In addition, the BCL parametrization of $f_+$ uses the known behaviour of its discontinuity
$\operatorname{Disc} f_+$
just above the pair-production threshold $(m_B + m_\pi)^2$, where one unit of orbital angular momentum imposes a
power-law $\operatorname{Disc}f_+ \propto p^{3/2}$ with $p$ the breakup momentum.
The absence of a $p^{1/2}$ term is then used to eliminate the expansion coefficient
$b^+_K$ in lieu of the coefficients $b^+_n$ with $n < K$.\\

There is some ambiguity as to how the scalar form factor $f_0$ should be parametrized,
which is not discussed in ref.~\cite{Bourrely:2008za}.
Commonly~\cite{Lattice:2015tia,Flynn:2015mha} $f_0$ is parametrized without the use of a pole,
due to the absence of a subthreshold bound state.
In addition, the behaviour of $\operatorname{Disc} f_0 \sim p^{1/2}$ just above the pair production threshold cannot be used to eliminate one of the
expansion coefficients. It is therefore ambiguous if $f_0$ should be expanded to order $K$ or $K - 1$
to ensure consistency when simultaneously fitting $f_+$.
In this section, we expand $f_0$ to order $z^{K-1}$ to make it compatible with the literature~\cite{Lattice:2015tia,Flynn:2015mha}.\\

For the tensor form factor $f_T$ most of the same considerations as for $f_+$ apply.
There is a single sub-threshold pole, which corresponds to the $B^*$ bound state. The factor $(1 - q^2/M_{B^*}^2)^{-1}$
accounts simultaneously for the asymptotic behaviours for $q^2 \to \infty$ and the bound state.
As for the vector form factor we expand to order $z^K$.
Above the pair production threshold, two units of orbital angular momentum impose that
$\operatorname{Disc}f_+ \propto p^{5/2}$. Hence, the absence of a $p^{1/2}$ term can again be used
to eliminate the expansion coefficient $b^T_K$ in lieu of the coefficients $b^T_n$ with $n < K$.

\begin{table}
    \def\arraystretch{1.3}
    \centering
    \begin{tabular}{ r r @{$^{+}_{-}$} l }
    \toprule
        \multicolumn{3}{c}{BCL parameters ($K=3$)} \\
        \hline
        $f_+(0)\,$ & $0.283$ & $^{0.027}_{0.027}$ \\
        $b^+_1\,$  & $- 1.0$ & $^{4.3}_{4.5}$     \\
        $b^+_2\,$  & $- 2.9$ & $^{6.2}_{5.8}$     \\
        \hline
        $b^0_1\,$ & $- 6.8$  & $^{6.3}_{6.9}$ \\
        $b^0_2\,$ & $4$      & $^{12}_{12}$   \\
        \hline
        $f_T(0)\,$ & $0.282$ & $^{0.026}_{0.026}$ \\
        $b^T_1\,$  & $- 0.7$ & $^{4.3}_{4.6}$     \\
        $b^T_2\,$  & $- 3.0$ & $^{6.3}_{5.9}$     \\
        \bottomrule
    \end{tabular}
    \label{tab:BCL-bfp-LCSR-only}
    \hspace{.05\textwidth}
    \begin{minipage}{.45\textwidth}
    \caption{
        The median values and central $68\%$ probability intervals obtained from the one-dimensional
        marginalized posterior distributions for the parameters of the common BCL parametrization \refeq{common-BCL} for the $K=3$ fit
        when fitted to the LCSR pseudo data points. The total $\chi^2$ is $0.017$ for $6$ degrees of freedom,
        corresponding to a $p$ value in excess of $99\%$ at the best-fit point.
    }
    \end{minipage}
\end{table}

Based on the above considerations, the common BCL parametrization then reads:
\begin{equation}
    \label{eq:common-BCL}
    \begin{split}
        f_+(q^2) & = \frac{f_+(q^2 = 0)}{1-q^2/m^2_{B^*}}\bigg[ 1 + \sum\displaylimits^{K-1}_{n=1} b_n^+\left (\bar{z}_n -(-1)^{n-K}\frac{n}{K}\bar{z}_K\right )\bigg],\\
        f_0(q^2) & = f_+(q^2 = 0)\bigg[1 + \sum\displaylimits_{n=1}^{K - 1}b_n^0 \bar{z}_n\bigg],\\
        f_T(q^2) & = \frac{f_T(q^2 = 0)}{1-q^2/m^2_{B^*}}\bigg[ 1 + \sum\displaylimits^{K-1}_{n=1} b_n^T\left(\bar{z}_n -(-1)^{n-K}\frac{n}{K}\bar{z}_K\right)\bigg],\\
    \end{split}
\end{equation}
with $\bar{z}_n \equiv z^n - z_0^n$, $z \equiv z(q^2; t_+, t_0)$, and $z_0 = z(0; t_+, t_0)$.
Here we manifestly fulfill the kinematical constraint $f_+(0) = f_0(0)$, which reduces the overall number of free parameters
by one.

\noindent We proceed to fit the common BCL parametrization \refeq{common-BCL} to the 14 LCSR pseudo data points
and their correlated uncertainties provided in \refsec{lcsr}. As discussed in that section, the correlated pseudo data points
can be counted as 14 independent observations.
Adding further data points is unlikely to increase the amount of information, due to the already
large degree of correlation among the data points.
In the fit to the LCSR prediction, the number of fit parameters is therefore limited to be smaller than 14,
corresponding to a maximal order $K=4$ in the $z$ expansion, which has eleven independent parameters.\\

We carry out two fits: one with $K=3$, and one with $K=4$.
Already for $K=3$ we obtain a good fit with $\chi^2/\mathrm{d.o.f.} \sim 0.017 / 6$ and a $p$ value in excess
of $99\%$. The goodness of fit therefore gives no indication that higher orders of $z$ are required in our fit model.
Nevertheless, we carry out a fit with $K=4$ to obtain
a handle on the systematic extrapolation error inherent to the form factor parametrization.
We show medians and central $68\%$ probability intervals of the marginalized one-dimensional posterior distributions
for each of the BCL parameters for the $K=3$ fit in \reftab{BCL-bfp-LCSR-only}.
Our corresponding results in the $K=4$ fit are compatible with the results of the $K=3$ fit at the $1\sigma$ level.
This is not surprising, since the uncertainty intervals for the shape parameters in the $K=4$ fit
are an order of magnitude larger than those in the $K=3$ fit, while the goodness of fit cannot be improved further.
We therefore use the $K=3$ fit as our default for numerical values and illustrations in this section.
Note that we do not use the unitarity bounds that have been formulated for exclusive $b\to u$ transitions form factors~\cite{Bourrely:2008za, Bharucha:2010im}. \\

\begin{figure}[p]
    \centering
    \includegraphics[width=0.47\textwidth]{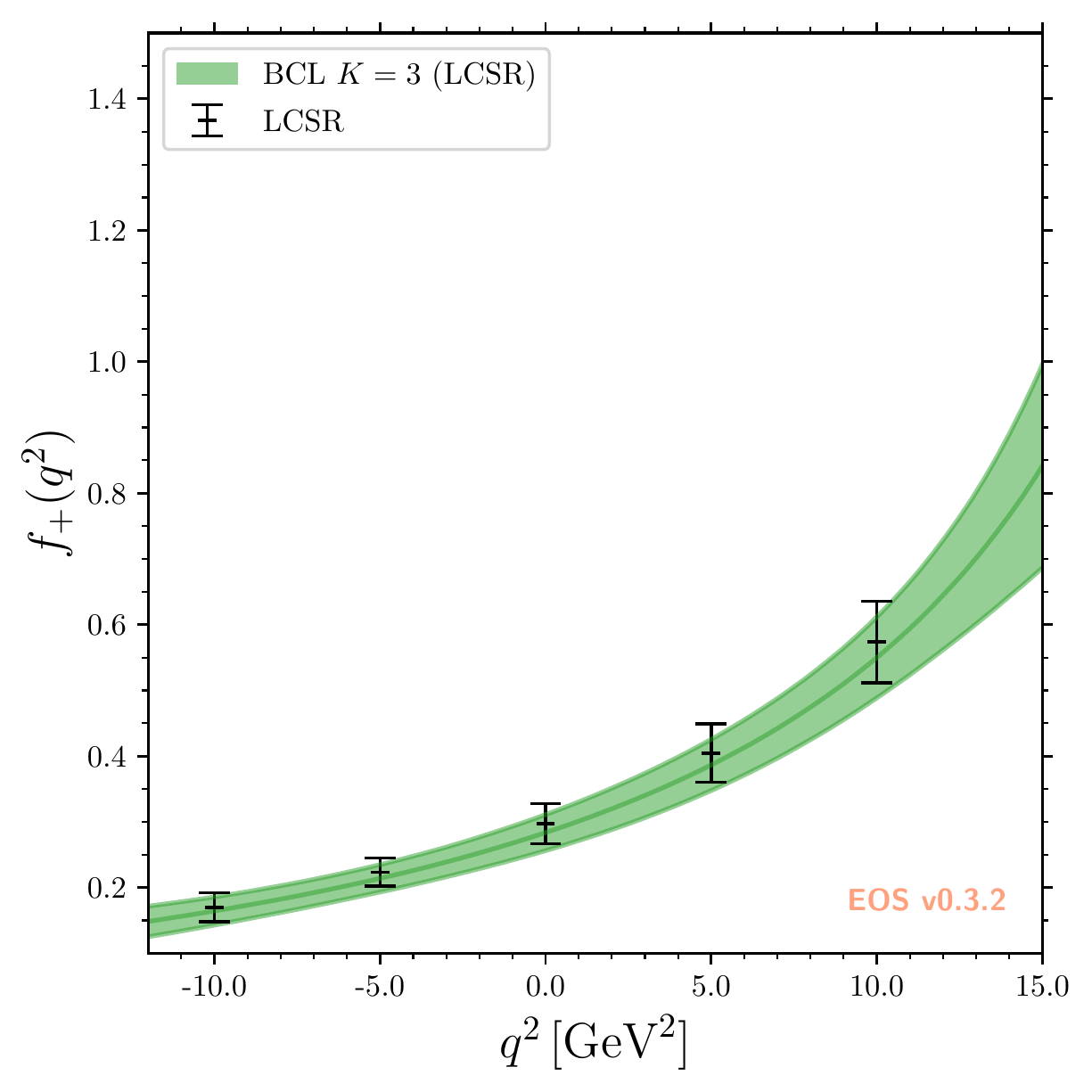}
    \includegraphics[width=0.47\textwidth]{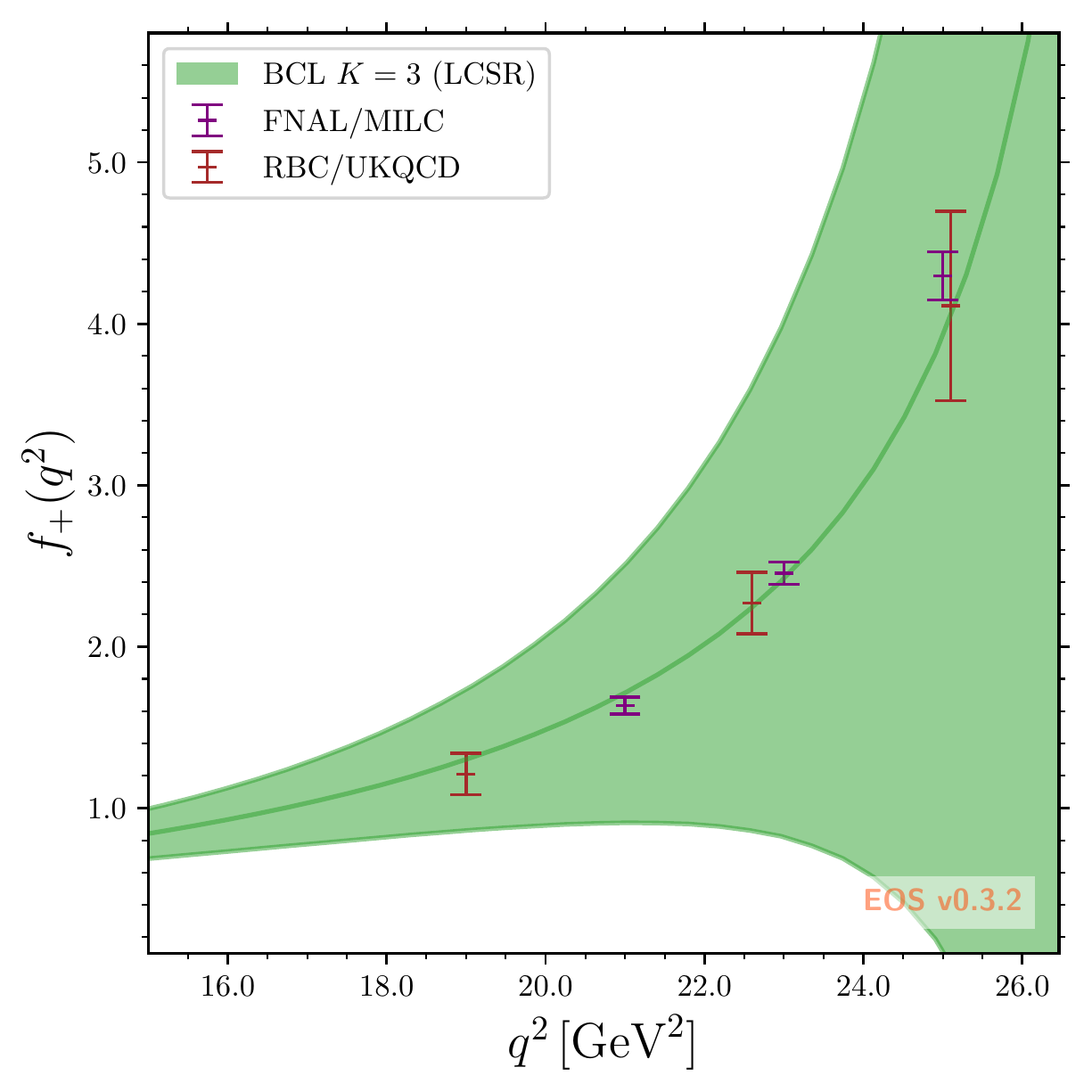}
    \includegraphics[width=0.47\textwidth]{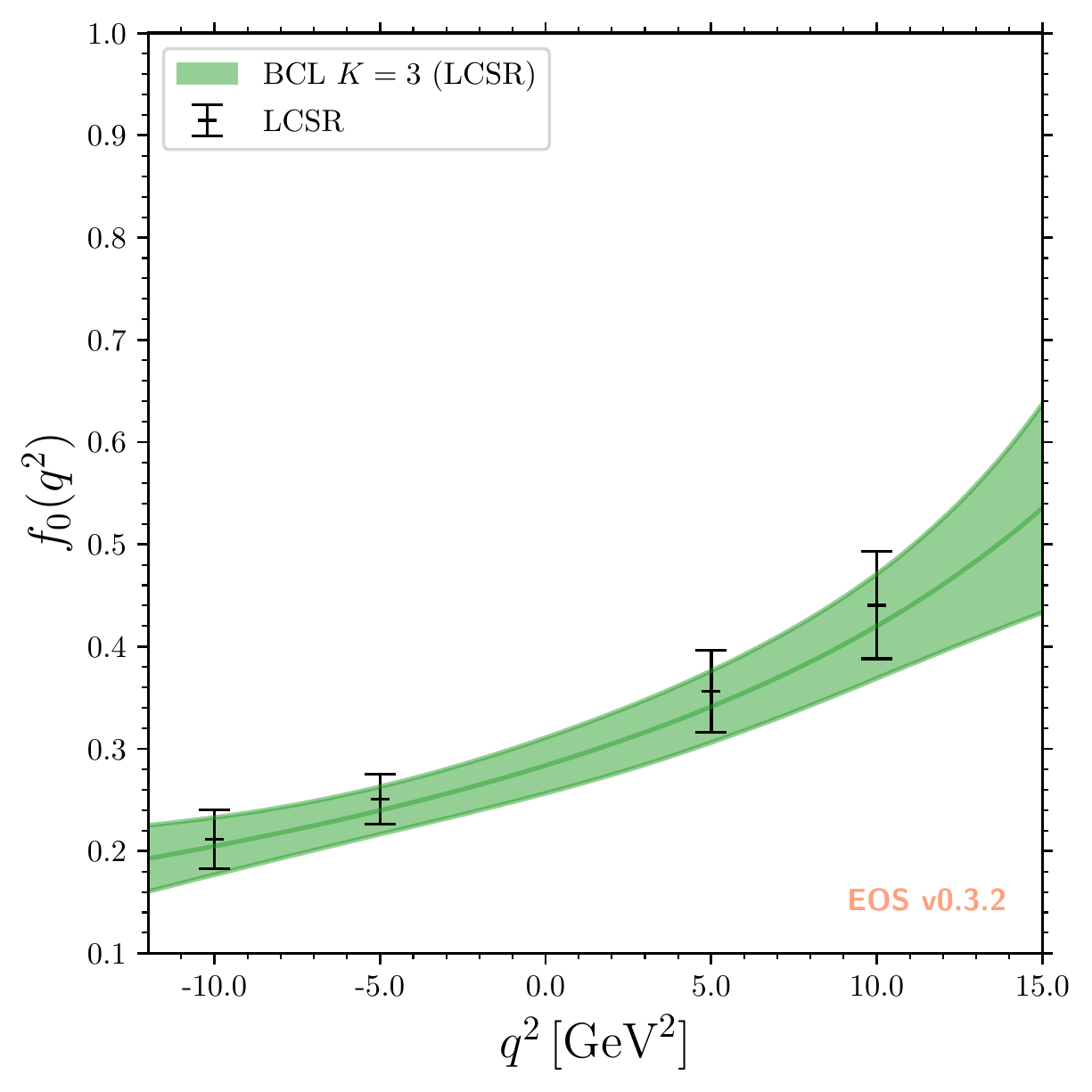}
    \includegraphics[width=0.47\textwidth]{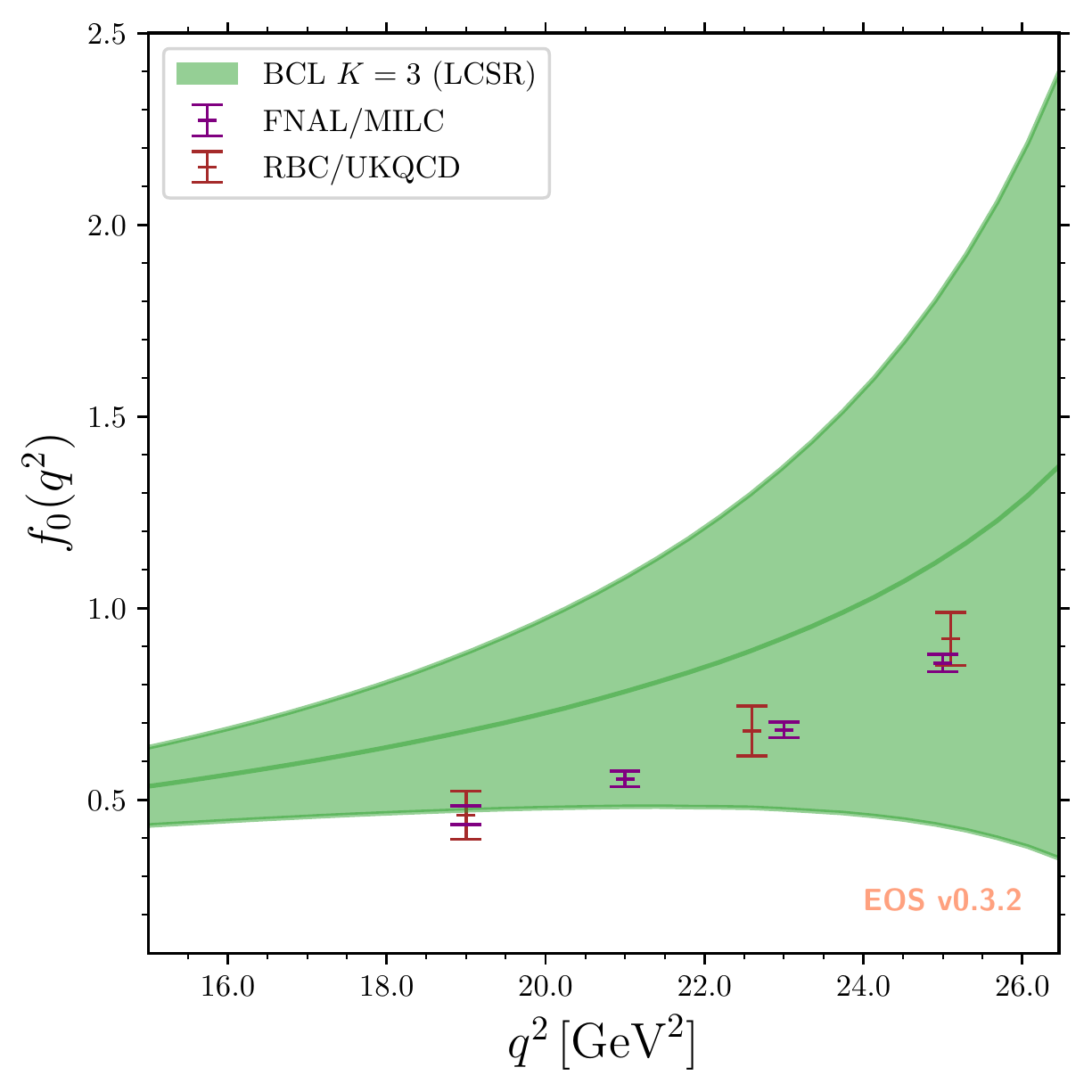}
    \includegraphics[width=0.47\textwidth]{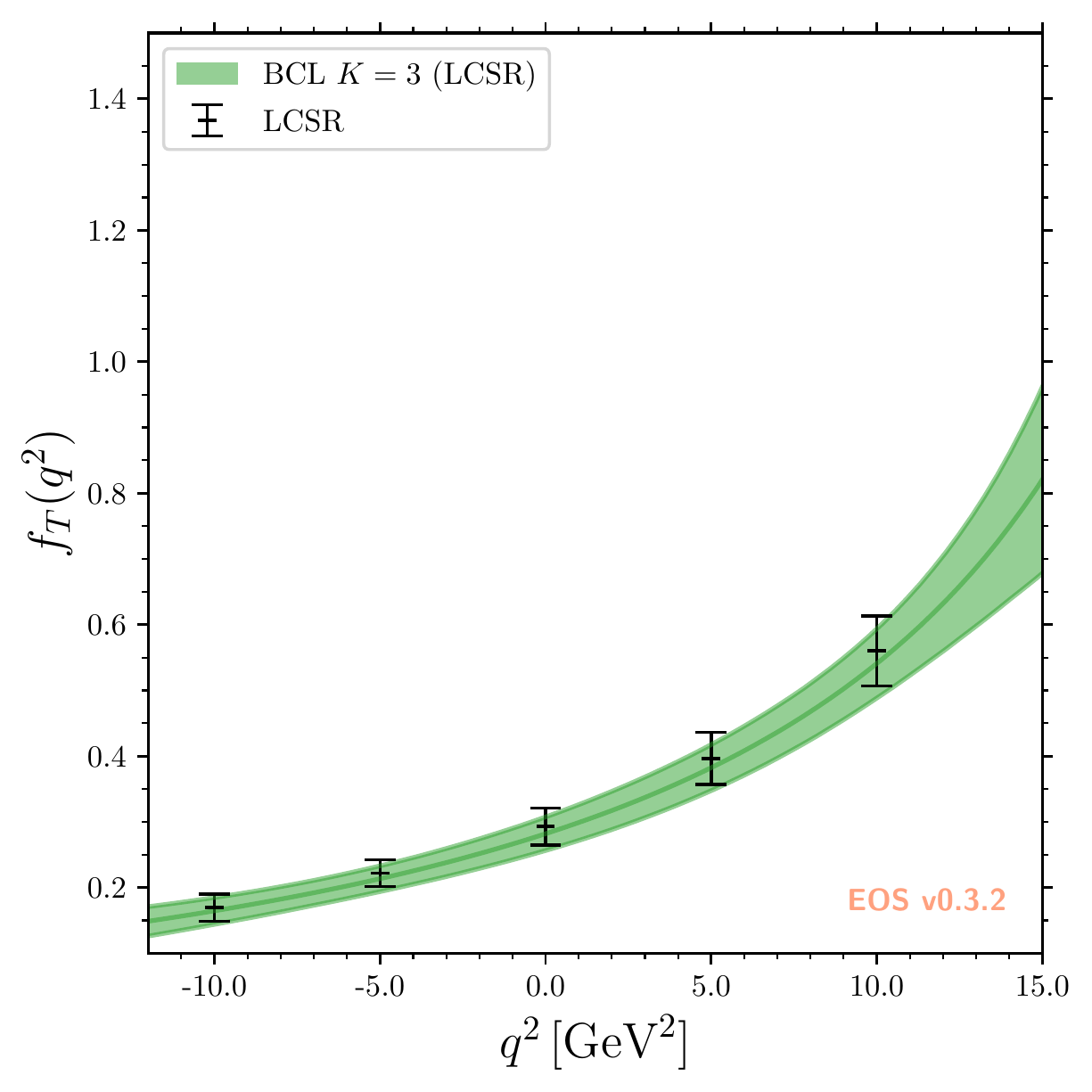}
    \includegraphics[width=0.47\textwidth]{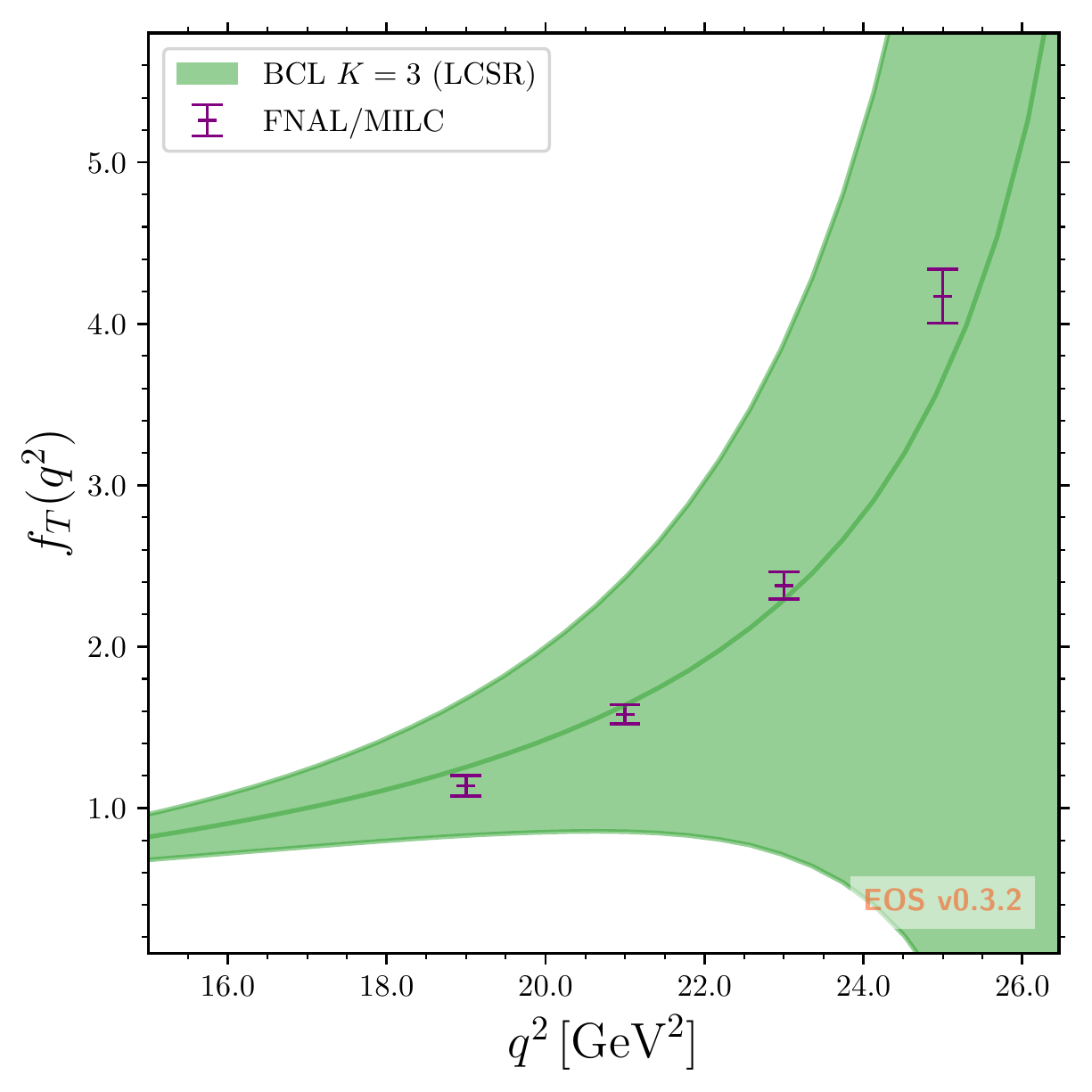}
    \caption{%
        Posterior-predictions for the form factors $f_+$ (top), $f_0$ (center), and $f_T$ (bottom)
        obtained from our fits of the common BCL parametrization (\ref{eq:common-BCL}) to \emph{only} the LCSR pseudo data points
        discussed in \refsec{lcsr}.
        Lattice QCD points are merely shown for illustrative purpose.
        The bands correspond to the envelope at $68\%$ probability.
    }
    \label{fig:BCL-fit-LCSR-only}
\end{figure}

We show our fit results in relation to the LCSR
pseudo data points in \reffig{BCL-fit-LCSR-only}.
This figure also indicates that our extrapolation to large $q^2$ has sizable uncertainties. Within these
uncertainties, our results are compatible with the available lattice QCD results~\cite{Lattice:2015tia, Bailey:2015nbd, Flynn:2015mha} for the
$\bar{B}\to \pi$ form factors. The latter are not part of the analyses in this section and are merely shown for an illustrative purpose.
In both the $K=3$ and the $K=4$ fit we observe that the bands of posterior-predictions at $68\%$ probability
do not correspond to the $68\%$ uncertainty regions of the data points. We find empirically that this effect
is caused by the large correlations among $q^2$-neighbouring data points and between the predictions for $f_+$ and $f_0$. The effect causes the BCL fit's uncertainty bands to
trail slightly below the form factor pseudo data points.\\

In the literature it is common to provide the $q^2$-integrated branching ratio in units of $|V_{ub}|^2$, based on LCSR results for the form factors. The integrated branching ratio can then be used to extract $|V_{ub}|$ from the experimental results, if the same integration range
is used.
We do not use this quantity in the phenomenological parts of our analysis, since we consider it less flexible than a full fit
to the form factor parameters and $|V_{ub}|$, which can also tell us about compatibility of the form factor shape between theory and experiment.
Stll, it can be useful for comparison with other works. We obtain:
\begin{equation}
    \frac{1}{|V_{ub}|^2} \int_0^{12\,\GeV^2} dq^2 \frac{d\mathcal{B}(\bar{B}^{0}\to \pi^{+} e^-\bar\nu_e)}{dq^2}
        =  7.6^{+1.6}_{-1.4}\,.
\end{equation}
The above value is compatible within uncertainties with the estimates given in the
literature~\cite{Khodjamirian:2011ub,Imsong:2014oqa,Khodjamirian:2017fxg}.

We can challenge our extrapolations of the LCSR results in several ways.\\
First, a dispersive representation
of $f_+(q^2)$ implies that:
\begin{equation}
    \label{eq:fp-residue-analytic}
    \operatorname{Res}_{q^2 \to m_{B^*}^2} f_+(q^2)
        = \frac{1}{2}f_{B^*} m_{B^*} g_{B^* B\pi}
        = 14.6 \pm 1.3 \, \GeV^2 \,, 
\end{equation}
for which we use $f_{B} = 190 \pm 1.3\,\MeV$ from a lattice calculation,
$f_B^*/f_B = 0.958 \pm 0.022$~\cite{Lubicz:2017asp} and $g_{B^*B\pi} = 30.1^{+2.6}_{-2.4}$ from a recent QCD light-cone sum rule
calculation~\cite{Khodjamirian:2020mlb}.
Since the common BCL parametrization for $f_+$ includes a pole for the $B^*$, we can obtain
an analytical formula for the residue. From our fit, we obtain
\begin{align}
    \operatorname{Res}_{q^2 \to m_{B^*}^2} f_+(q^2) = 12\pm 29 \,  \GeV^2 \,,
\end{align}
which is in agreement with \refeq{fp-residue-analytic} within its sizable uncertainties. The uncertainties presented
are of parametric origin only. Systematic uncertainties due to higher orders
in the $z$ expansion are not taken into account, and could be sizable due to the magnitude of $z(q^2 = m_{B^*}^2)$.
The residue is therefore not immediately useful to check the validity of our extrapolation to large $q^2$.

Second, the soft-pion theorem
relates the form factor $f_0$ to the $B$-meson and pion decay
constants~\cite{Isgur:1989qw,Dominguez:1990mi,Wise:1992hn,Burdman:1992gh,Wolfenstein:1992xh} as 
\begin{align}
    f_0(t_-) + f_0(t_+)
        & = \frac{2 f_B}{f_\pi} \left[1 - \frac{m_u + m_d}{m_d + m_b}\right] = 2.914 \pm 0.092\,.
         \label{eq:f0soft}
\end{align}
The relation holds even at next-to-leading order in $1/m_b$, and including short-distance corrections \cite{Burdman:1993es}.
Here we use the same numerical inputs as above and additionally $f_{\pi} = 130.2 \pm 0.8 \MeV$~\cite{Zyla:2020zbs}.
Our fit of the form factors gives us
\begin{align}
    \label{eq:Callan-Treiman-type-rel}
    f_0(t_-) + f_0(t_+)\bigg|_{\text{LCSR only}}
        & = 5.1\pm 6.3\,.
\end{align}
Although our results are consistent with the expectation in \refeq{f0soft}, the uncertainties
are so large that we cannot use the above relations to carry out a meaningful test of the validity
of our extrapolation.

\begin{figure}[t]
    \centering
    \begin{minipage}[c]{.68\textwidth}
        \includegraphics[width=\textwidth]{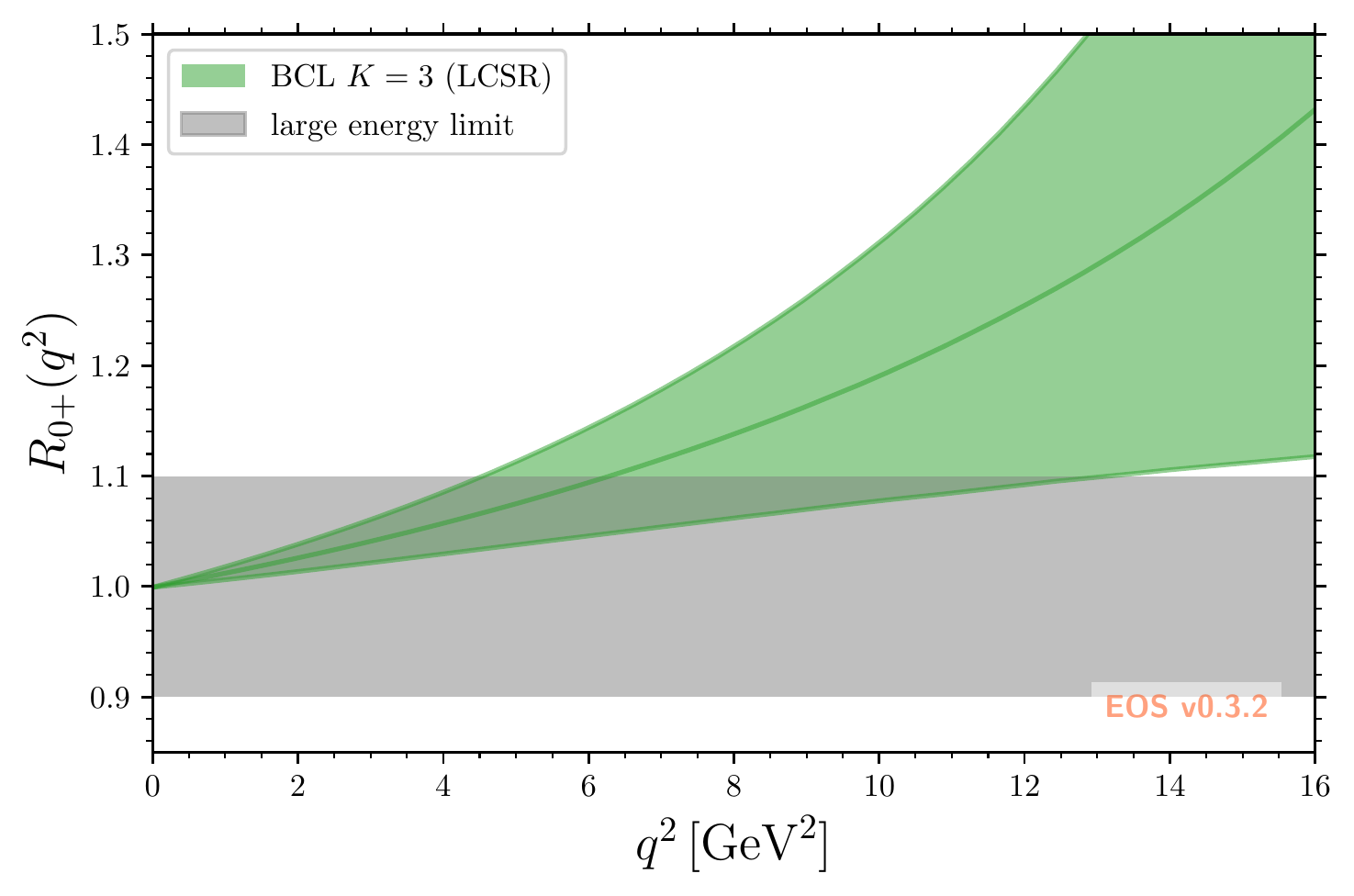}
    \end{minipage}
    \hspace{.02\textwidth}
    \caption{
    The ratio of form factors $R_{0+}(q^2)$ as defined in the text.
    We extrapolate based on the fit to LCSR pseudo data
    and compare with the symmetry limit (\ref{eq:HQ_lim_f0}), which is valid
    for large $\pi$ energy in the $B$-meson rest frame or equivalently at small $q^2$.
    }
    \label{fig:le-ratio-LCSR}
\end{figure}

Third, in the large-energy symmetry limit the form factors $f_+$ and $f_0$ are related via~\cite{Beneke:2000wa}: 
\begin{equation}
    \label{eq:HQ_lim_f0}
    f_0(q^2)  = \frac{m_B^2 + m_{\pi}^2 - q^2}{m_B^2} f_+(q^2) + \mathcal{O}\left(\frac{\Lambda_\text{had}}{E_\pi},\frac{\Lambda_\text{had}}{m_B}\right)\,.
\end{equation}
Here $E_\pi$ is the energy of the $\pi$ in the $B$ rest frame.
A useful measure of compatibility can therefore be obtained through the ratio
\begin{equation}
    R_{0+}(q^2) = \frac{m_B^2}{m_B^2+m_{\pi}^2-q^2}\frac{f_0(q^2)}{f_+(q^2)}.
\end{equation}
We show this ratio in \reffig{le-ratio-LCSR} based on an extrapolation of our LCSR form factors.
We find that the LCSR results are consistent with the large-energy limit within $\sim 10\%$
uncertainty up to $\simeq 13\,\GeV^2$, \emph{i.e.} within the whole region of applicability of the LCSRs.\\

In the next section we proceed with form factor extractions in a combined fit to LCSR and lattice QCD inputs.
This procedure further constrains the form factors at high $q^2$ and reduces the uncertainties appreciably.

\section{Interpolation between LCSR and lattice QCD results}
\label{sec:interpolation}

In this section we proceed to challenge the LCSR results obtained in \refsec{lcsr} and their extrapolation
to large $q^2$ in \refsec{extrapolation} with precise results for the $\bar{B}\to \pi$ form factors
obtained from lattice QCD simulations. These lattice QCD results exhibit very small uncertainties for 
$19\,\GeV^2 \lesssim q^2 \lesssim 25\,\GeV^2$, outside the reach of the light-cone sum rules.
In this work we use two independent sets of lattice QCD results. The first set is provided by the FNAL/MILC
collaboration~\cite{Lattice:2015tia, Bailey:2015nbd} based on $N_f=2+1$ gauge ensembles and
a staggered-quark action.
The second set is provided by the RBC/UKQCD collaboration~\cite{Flynn:2015mha} based on $N_f=2+1$ gauge
ensembles with domain wall fermions.\\

We refrain from using information from an older analysis by the HPQCD collaboration~\cite{Dalgic:2006dt} for
two reasons: first, it shares some of the $N_f=2+1$ ensembles with the results published by FNAL/MILC~\cite{Lattice:2015tia};
second, it does not provide correlations between the $f_+$ and $f_0$ results.
We also refrain from using a more recent HPQCD analysis~\cite{Colquhoun:2015mfa} providing a single, very precise value for
$f_0$ at zero-recoil. Again, some of the $N_f=2+1$ ensembles are shared with the FNAL/MILC analysis, and
we cannot account for the correlations between the HPQCD and the FNAL/MILC results.\\

The usage of RBC/UKCQCD data is straightforward, since ref.~\cite{Flynn:2015mha} provides
both the $f_+$ and the $f_0$ form factor in three different $q^2$ points including their correlations.
We therefore include these data in our likelihood as a multivariate gaussian constraint.\\
The usage of the FNAL/MILC data is more involved, since refs.~\cite{Lattice:2015tia, Bailey:2015nbd}
do not provide data points for any of the form factors. Instead, these references provide the outcome
of a BCL fit to the data points. As discussed below, we see the need to modify the BCL parametrization,
making it impossible to use the BCL results of FNAL/MILC collaboration as is.
Instead, we are forced to use the BCL results to produce pseudo data points of the form factors.
We produce three such points for $f_+$ and four points both for $f_0$ and for $f_T$. The points are
chosen in the range $19\,\GeV^2 \leq q^2 \leq 25\,\GeV^2$. Based on information provided
in ref.~\cite{Lattice:2015tia, Bailey:2015nbd}, we have chosen this range of $q^2$ to minimize the total uncertainty.
The smaller number of points for $f_+$ is due to a peculiarity in the BCL fit results. We find that
the covariance matrix provided in ref.~\cite{Lattice:2015tia} is singular. This can be understood,
since in that work the identity $f_+(0) = f_0(0)$ is not manifestly fulfilled by the parametrization.
Based on the authors' suggestions~\cite{Bazavov:2019aom}, we replace the coefficient $b_3^+$ (in their notation)
by a linear combination of the remaining coefficients, such that the identity $f_+(0)=f_0(0)$ is manifestly
fulfilled. This replacements requires the removal of the row and column associated with $b_3^+$ from the correlation matrix,
reducing the number of free parameters to three. Hence, the maximal number of independent pseudo data points
from FNAL/MILC is now also limited to three.\\

An overview of the data points used is provided in \reftab{fit_theory_input}.\\

\begin{table}[t]
    \centering
    \begin{tabular}{ c  c  c  c  c }
        \toprule
        form factor            & \# of points
                                     & $q^2$ values (in $\mathrm{GeV}^2$) & type & source                           \\
        \hline
        \multirow{3}{*}{$f_+$} & $5$ & $-10.0$, $-5.0$, $0.0$, $5.0$, $10.0$   & LCSR & this work                        \\
                               & $3$ & $21.0$, $23.0$, $25.0$                  & LQCD & FNAL/MILC \cite{Lattice:2015tia} \\
                               & $3$ & $19.0$, $22.6$, $25.1$                  & LQCD & RBC/UKQCD \cite{Flynn:2015mha}   \\
        \hline
        \multirow{3}{*}{$f_0$} & $4$ & $-10.0$, $-5.0$, $5.0$, $10.0$          & LCSR & this work                        \\
                               & $4$ & $19.0$, $21.0$, $23.0$, $25.0$          & LQCD & FNAL/MILC \cite{Lattice:2015tia} \\
                               & $3$ & $19.0$, $22.6$, $25.1$                  & LQCD & RBC/UKQCD \cite{Flynn:2015mha}   \\
        \hline
        \multirow{2}{*}{$f_T$} & $5$ & $-10.0$, $-5.0$, $0.0$, $5.0$, $10.0$   & LCSR & this work                        \\
                               & $4$ & $19.0$, $21.0$, $23.0$, $25.0$          & LQCD & FNAL/MILC \cite{Bailey:2015nbd}  \\
        \bottomrule
    \end{tabular}
    \label{tab:fit_theory_input}
    \caption{
    A complete list of the data points for the 3 transition form factors used in the combined fit.
    }
\end{table}

With the likelihood for the lattice QCD results at hand, we carry out a simultaneous fit of the common BCL parametrization
in \refeq{common-BCL} to both the LCSR pseudo data points and the lattice QCD data points.
We find that for $K=3$ the fit yields a minimal $\chi^2 \simeq 154$. Given $23$
degrees of freedom in the fit, this corresponds to a $p$ value considerably smaller than our a-priori
threshold of $3\%$. We therefore have to reject this fit.
Investigating the BCL paramatrization with $K=4$, we find better agreement with a minimal $\chi^2 = 27.5$
for $20$ degrees of freedom. The corresponding $p$ value of $12\%$ is acceptable.
For both cases, a visual comparison of the extrapolation of the LCSR results for $f_+$ and $f_T$ with the lattice data
as shown in \reffig{BCL-fit-LCSR-only} does not give any reason to expect a bad fit. However, the same figure
illustrates that the extrapolation of $f_0$ is not easily compatible with the lattice points.
We therefore conclude that the goodness of fit of the overall analysis hinges crucially on the
correlations between $f_0$ and the other form factors.\\

The surprising result for the fit with the common BCL parametrization leads us to investigate alternative
fit models. We modify the parametrization of the $f_0$ form factor by including a pole above the $B\pi$ pair production threshold, corresponding to a scalar $B\pi$ resonance. 
No such resonance has been observed
yet. Hence, we have to rely on models for a prediction of its mass $m_{B_0}$.
In the literature a wide range of values from different models for $m_{B_0}$ can be found:
$m_{B_0} \in [5.526, 5.756]\,\mathrm{GeV}$~\cite{Cheng:2017oqh}.
Here, we use $m_{B_0} = 5.54\,\GeV$, which is compatible with other form factor parametrisations
involving a scalar resonance within the EOS software, ensuring their interoperability with ours.
However, we emphasize that the position of pole above the $B\pi$ threshold suffices to improve the fit quality
dramatically. We have explicitly checked that varying the mass does not influence our results qualitatively:
Varying the value of $m_{B_0}$ in the aforementioned range is always compensated by minor shifts to the central
values of the remaining parameters.
The modified BCL parametrization then reads:
\begin{equation}
\label{eq:modified-BCL}
\begin{aligned}
    f_0(q^2)
        & = \frac{f_+(z_0)}{1-q(z)^2/m^2_{B_0}}\bigg[1 + \sum\displaylimits_{n=1}^{K}b_n^0 \bar{z}_n\bigg],\\
    f_+(q^2),\, f_T(q^2)
        & : \text{unchanged with respect to eq.~\eqref{eq:common-BCL}}\,,
\end{aligned}
\end{equation}
where we also increase the maximal order of the $z$ expansion for the $f_0$ form factor. In this way
we now use the same number of shape parameters for each form factor. (Note that for
$f_+$ and $f_T$ one shape parameter is fixed as discussed in \refsec{extrapolation}. 
We further note that our modification does not allow to apply the unitarity bounds for the $f_0$ form factor as is.
However, alternative parametrizations such as the BGL parametrization can account for above-threshold
poles in the formulation of the unitarity bounds~\cite{Caprini:2017ins}, which we do not consider here.
We repeat the fits with the modified BCL parametrization in \refeq{modified-BCL} with $K=3$ and $K=4$.
In both cases we obtain acceptable to good fits, with $p$ values of $52\%$ and $54\%$ respectively.\footnote{
The inclusion of the scalar resonance makes up for the majority of the $p$ value improvement.
Keeping only the pole and fixing the additional shape parameter to zero yields $p$ values of
$7\%$ and $60\%$, respectively. 
}
Effectively, the pole modifies the shape parameters and implicitly allows for more flexibility of the fit.
Explicitly expanding the $B^*$ pole factor in $z$ around $q^2=0$ yields:
\begin{equation}
    \frac{1}{1-q(z)^2/m_{B^*}^2}
        \approx \frac{1}{1-\frac{t_0}{m_{B^*}^2}} + 4 \frac{m_{B^{*}}^2(t_0 - t_+)}{(m_{B^*}^2 - t_0)^2} z
            + \mathcal{O}\left(z^2\right)\,.
\end{equation}
This illustrates that additional powers of $z$ are now available to relieve the apparent tension between
the LCSR and LQCD data of $f_0$.\\

The median values and central $68\%$ probability intervals for each marginalized one-dimension
posteriors are provided in \reftab{BCL-bfp-LCSR+lQCD}.
The covariance matrix is provided as an ancillary file together with the arXiv preprint of this paper.
We find that the fit parameter values for $K=4$ are consistent with the ones for $K=3$ within uncertainties. 
We also investigate the $K=5$ case for which the $p$ value increases insignificantly compared to the $K=4$ case.
Although the $K=3$ fit is also acceptable, we consider the $K=4$ fit to be our main result. The reason is
that at $K=4$ the fit can account for an additional systematic uncertainty due higher orders in the
$z$ expansion.\\

\begin{figure}[p]
    \centering
    \includegraphics[width=0.47\textwidth]{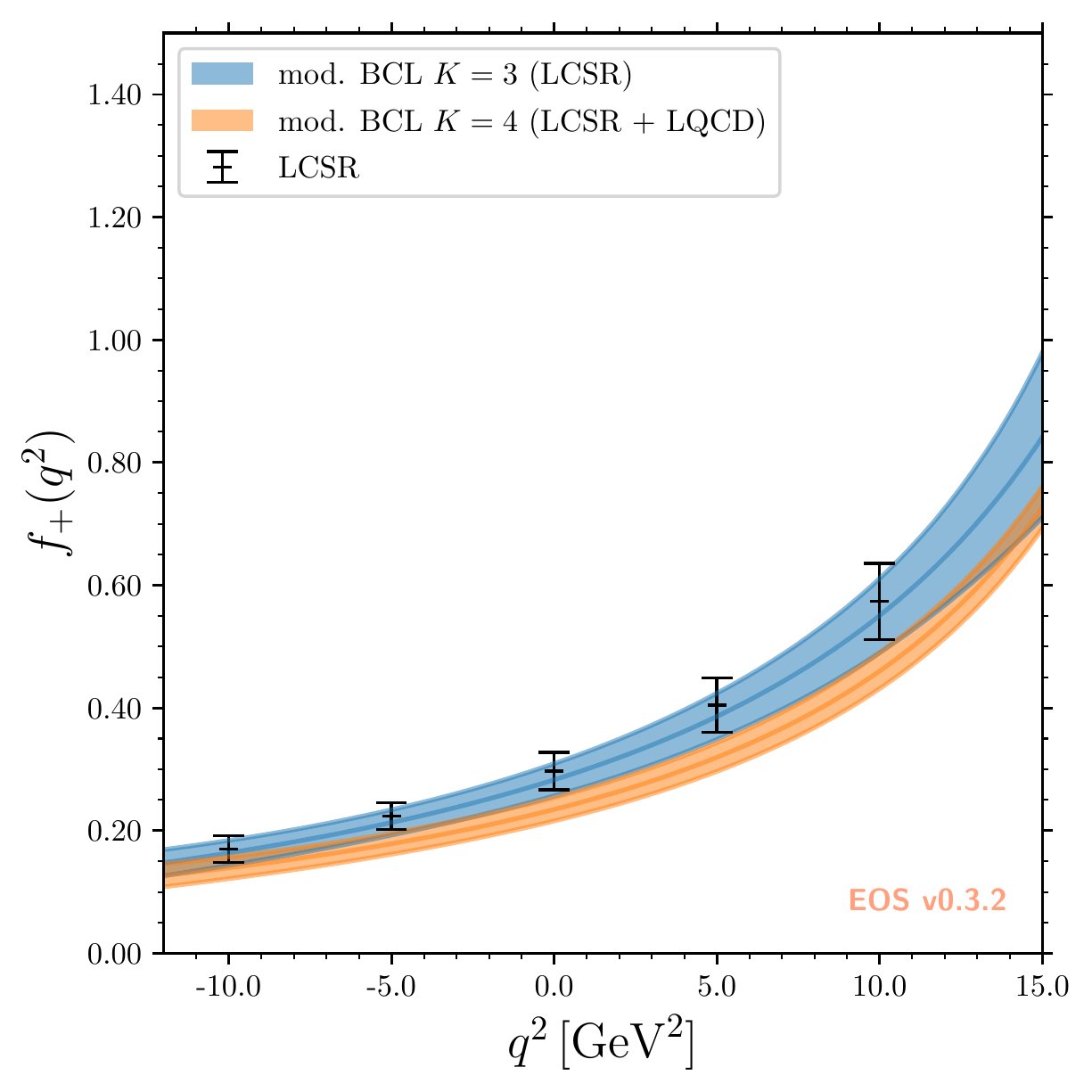}
    \includegraphics[width=0.47\textwidth]{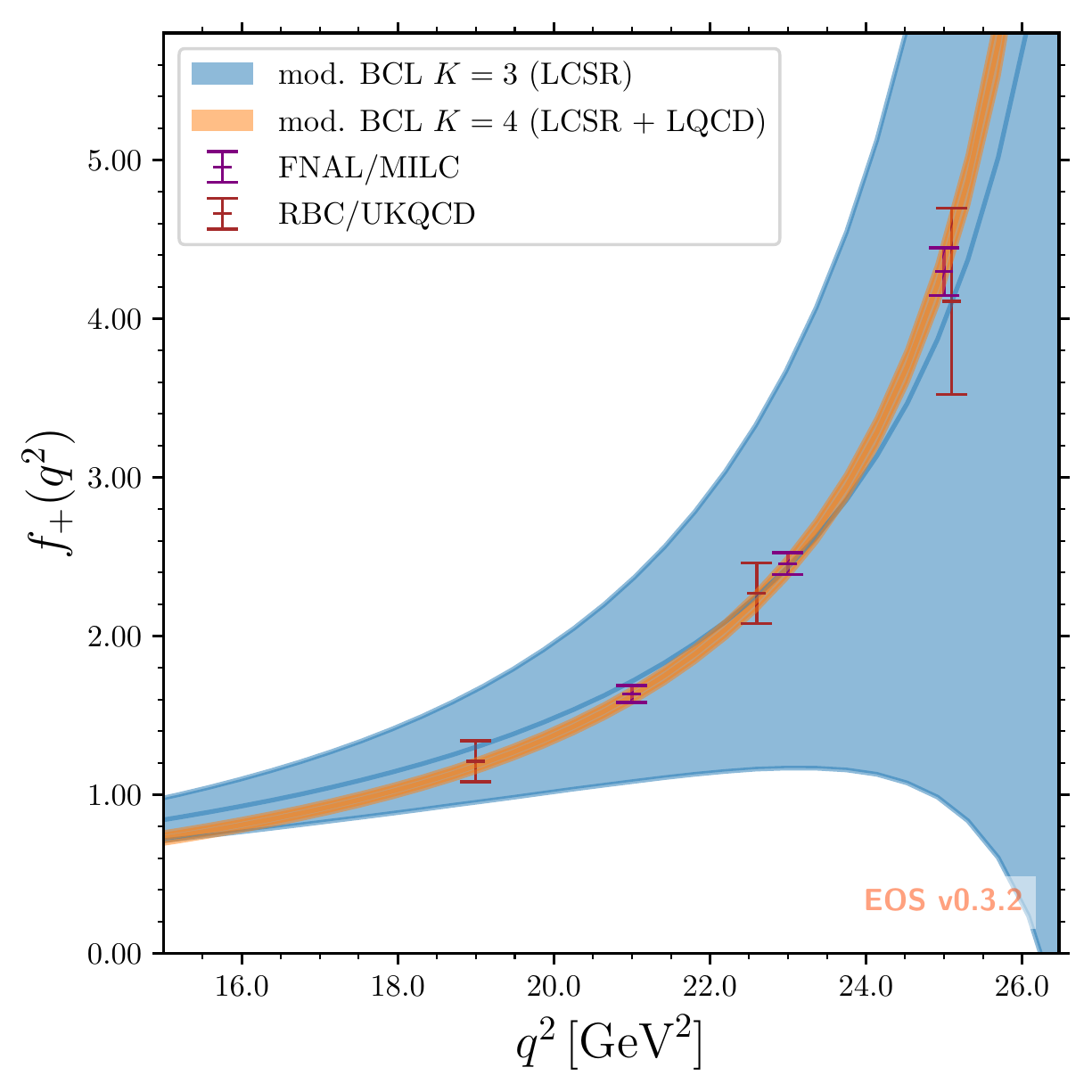}
    \includegraphics[width=0.47\textwidth]{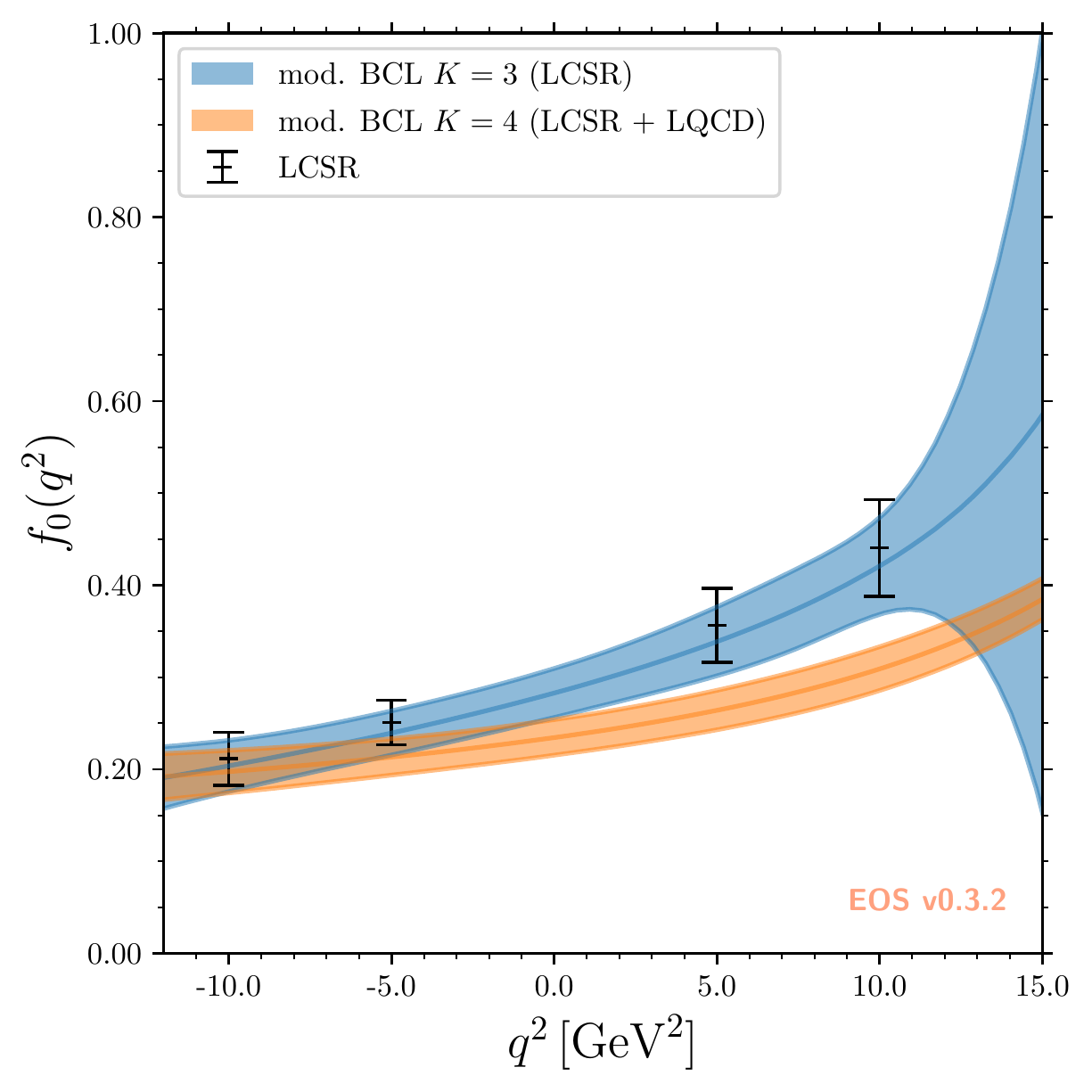}
    \includegraphics[width=0.47\textwidth]{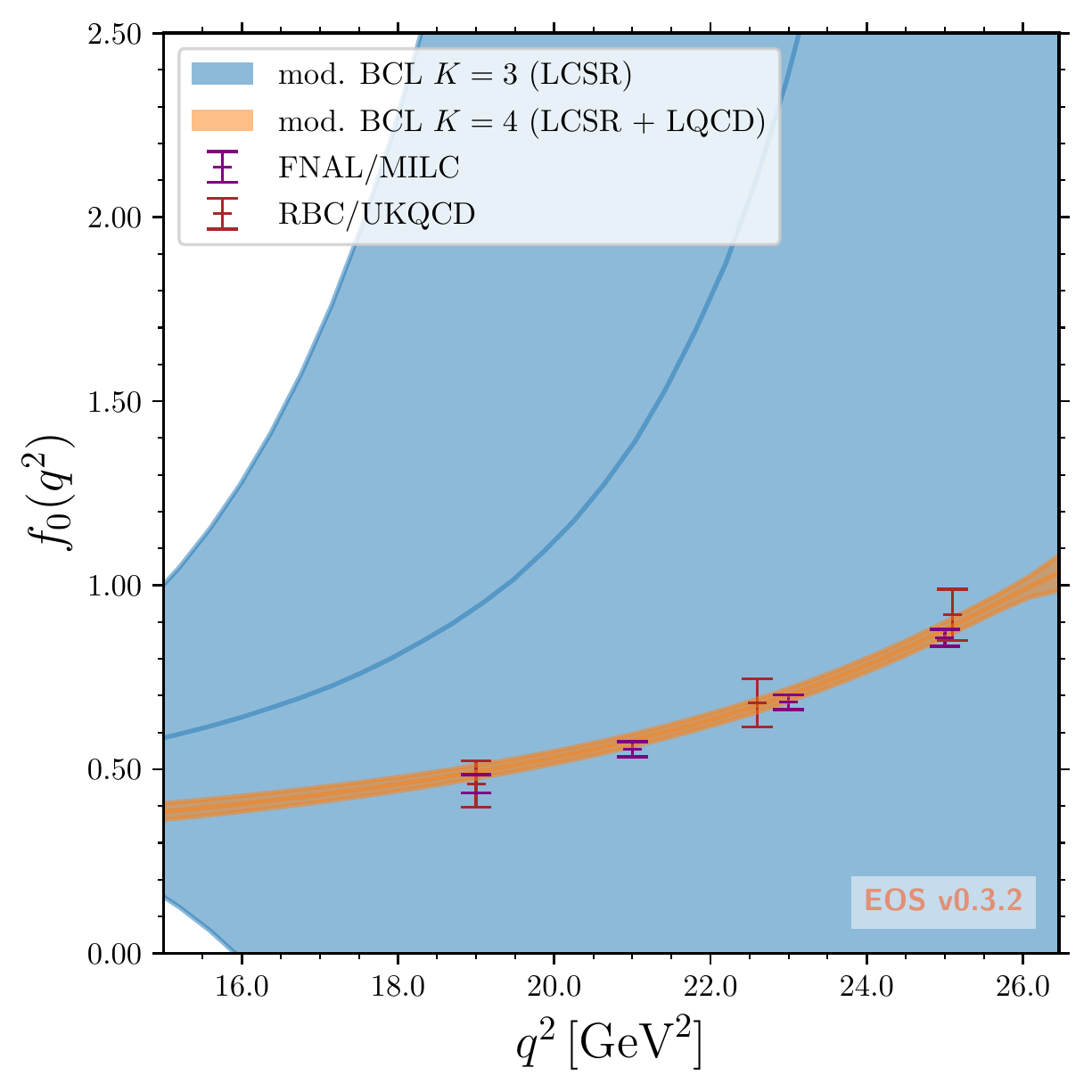}
    \includegraphics[width=0.47\textwidth]{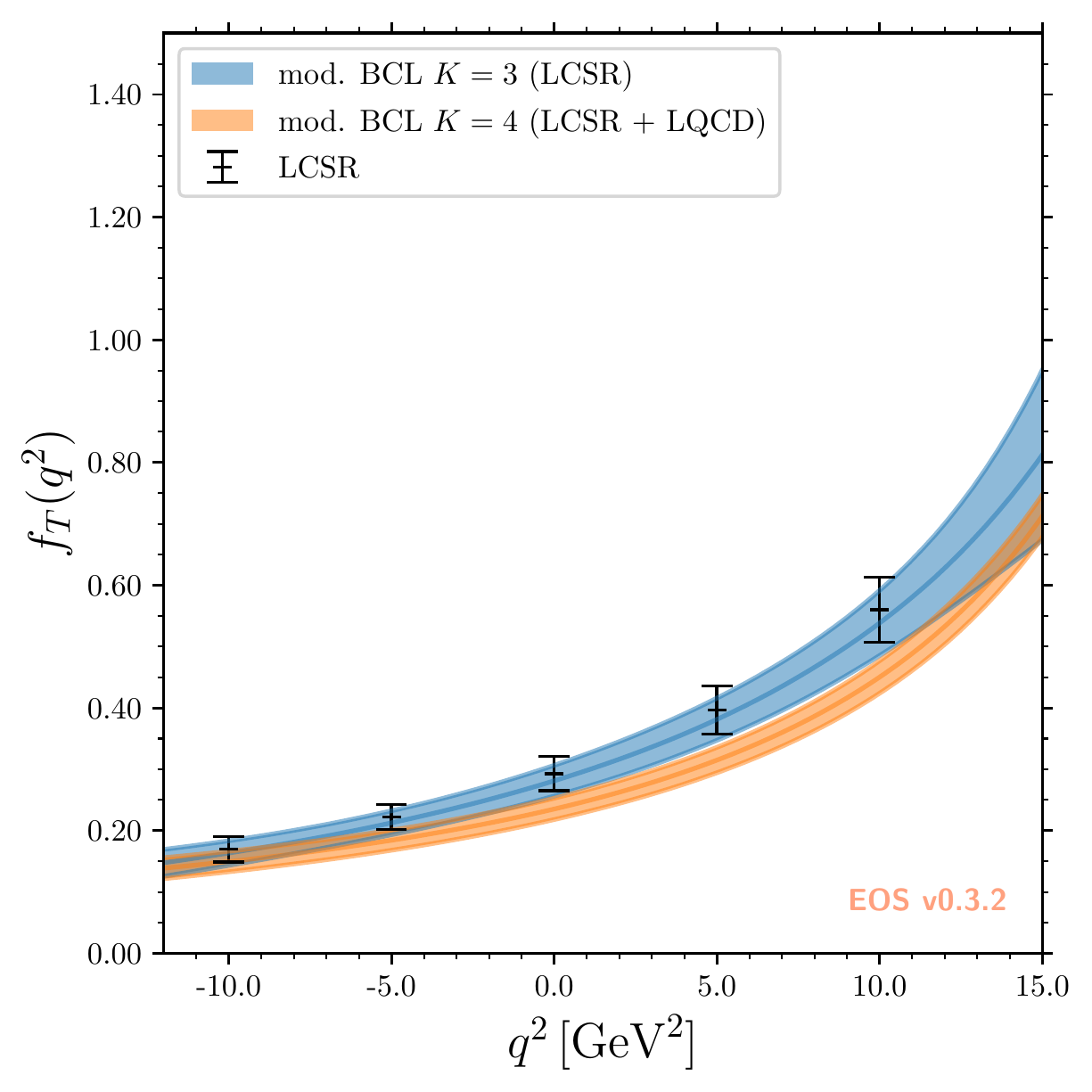}
    \includegraphics[width=0.47\textwidth]{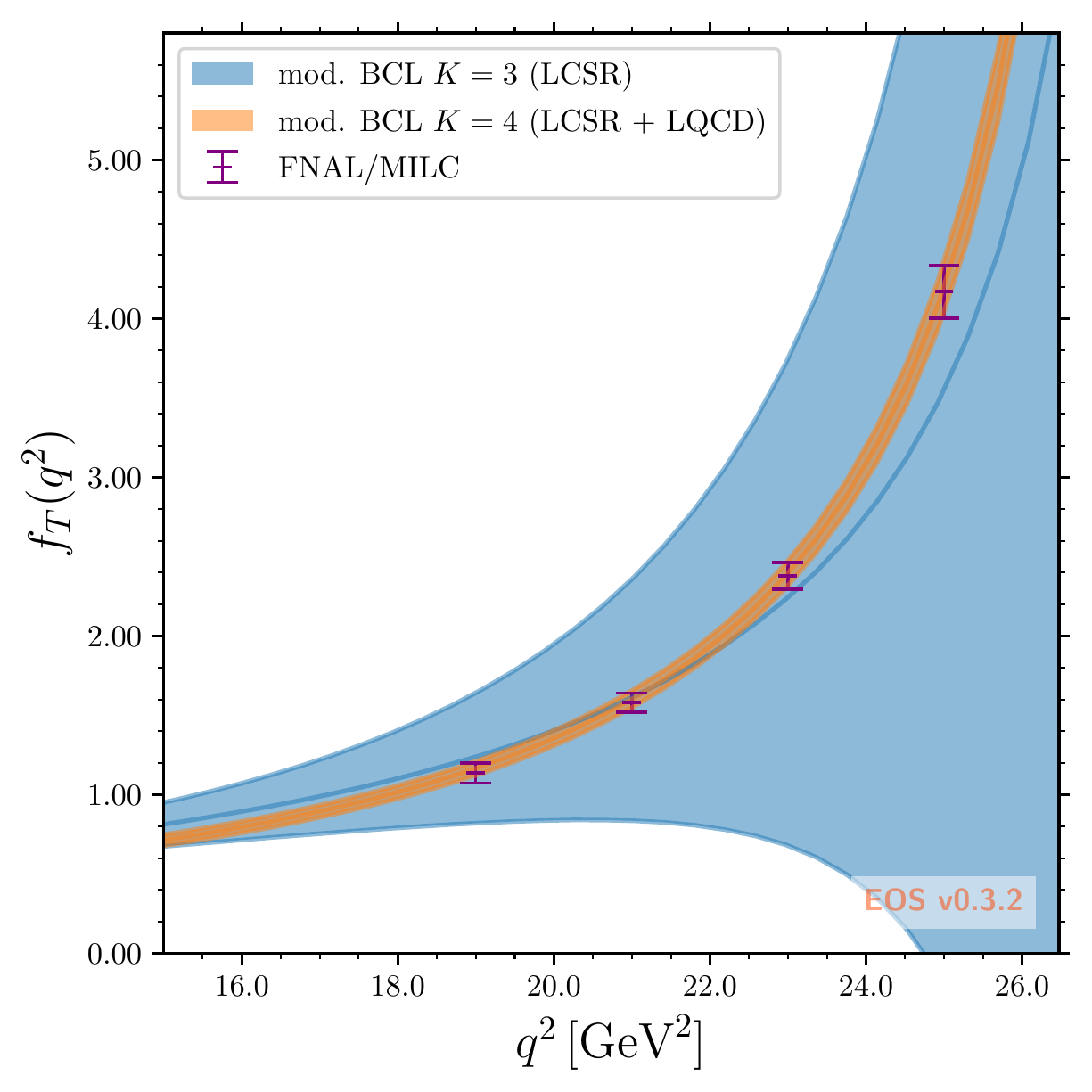}
    \caption{
        Posterior-predictions for the form factors $f_+$ (top), $f_0$ (center), and $f_T$ (bottom)
        obtained from our fits of the modified BCL parametrization (\ref{eq:modified-BCL}) to the LCSR pseudo data points only (blue bands),
        and both the LCSR and lattice QCD inputs (orange bands).
        The bands correspond to the envelope at $68\%$ probability.
        }
    \label{fig:modified-BCL-fit-LCSR+LQCD}
\end{figure}

Plots of the posterior predictions for each form factor are provided in \reffig{modified-BCL-fit-LCSR+LQCD}.
A cursory glance at the $f_+(0)$ plot suggest a $\sim1.4\sigma$ deviation between
the fit to the LCSR results only and the fit to the combined LCSR+LQCD likelihood.
We remind the reader that, in addition to the normalization $f_+(q^2 = 0)$, the fits also need to bring
LCSR and LQCD predictions for the slope into mutual agreement.
Hence, we emphasize that
a naive interpretation \emph{is not useful}, due to strong correlation between the normalization and the shape parameters.
As a consequence, we cannot accurately
compute the compatibility of only the normalization $f_+(q^2 = 0)$, and the overall goodness-of-fit diagnostics of our fits, such as the
$p$ value, must suffice.

\begin{table}[t]
    \def\arraystretch{1.3}
    \centering
    \begin{tabular}{ c r @{$^{+}_{-}$} l r @{$^{+}_{-}$} l r @{$^{+}_{-}$} l }
        \toprule
        \multirow{2}{*}{\diagbox[height=2\line]{param.}{scenario}} & \multicolumn{4}{c}{LCSR+LQCD} & \multicolumn{2}{c}{LCSR}  \\ 
                            & \multicolumn{2}{c}{$K=3$} & \multicolumn{2}{c}{$K=4$}     & \multicolumn{2}{c}{$K=3$} \\
        \hline
        $f_+(0)$ &  $  0.237 $ & $^{0.017}_{0.017}$ & $  0.235 $ & $^{0.019}_{0.019}$  & $  0.283 $ & $^{0.027}_{0.027}$ \\
        $b^+_1$  &  $- 2.38 $  & $^{0.33}_{0.38}$   & $- 2.45 $  & $^{0.49}_{0.54}$    & $- 1.0 $   & $^{3.5}_{3.6}$ \\
        $b^+_2$  &  $- 0.82 $  & $^{0.76}_{0.81}$   & $- 0.2 $   & $^{1.1}_{1.2}$      & $- 2.8 $   & $^{4.9}_{4.7}$ \\
        $b^+_3$  &\multicolumn{2}{c}{---}           & $- 0.9 $   & $^{4.2}_{4.0}$      & \multicolumn{2}{c}{---} \\
        $b^0_1$  &  $  0.48 $ & $^{0.07}_{0.07}$    & $  0.40 $  & $^{0.18}_{0.20}$    & $- 5 $     & $^{52}_{51}$ \\
        $b^0_2$  &  $  0.14 $ & $^{0.39}_{0.44}$    & $  0.1 $   & $^{1.1}_{1.2}$      & $ 22 $     & $^{200}_{200}$ \\
        $b^0_3$  &  $  2.79 $ & $^{0.71}_{0.77}$    & $  3.7 $   & $^{1.6}_{1.6}$      & $- 32$     & $^{240}_{240}$ \\
        $b^0_4$  &\multicolumn{2}{c}{---}           & $ 1 $      & $^{14}_{13}$        & \multicolumn{2}{c}{---} \\
        $f_T(0)$ &  $  0.240 $ & $^{0.016}_{0.016}$ & $  0.235 $ & $^{0.017}_{0.017}$  & $  0.281 $ & $^{0.025}_{0.025}$ \\
        $b^T_1$  &  $- 2.05 $  & $^{0.32}_{0.36}$   & $- 2.45 $  & $^{0.45}_{0.50}$    & $- 0.6 $   & $^{4.2}_{4.4}$ \\
        $b^T_2$  &  $- 1.45 $  & $^{0.63}_{0.66}$   & $- 1.08 $  & $^{0.68}_{0.71}$    & $- 3.2 $   & $^{5.9}_{5.8}$ \\
        $b^T_3$  &\multicolumn{2}{c}{---}           & $  2.6 $   & $^{2.1}_{2.0}$      & \multicolumn{2}{c}{---} \\
        \hline
        $p$ value               & \multicolumn{2}{c}{$\sim 52\%$}     & \multicolumn{2}{c}{$\sim 54\%$}     & \multicolumn{2}{c}{$\sim 100\%$} \\
        $\chi^2/\mathrm{d.o.f}$ & \multicolumn{2}{c}{$\sim 21.01/22$} & \multicolumn{2}{c}{$\sim 17.75/19$} & \multicolumn{2}{c}{$\sim 0.0278/5$} \\
        \bottomrule
    \end{tabular}
    \caption{The median values and central $68\%$ probability intervals for the parameters of the modified
        BCL parametrisation from~\refeq{modified-BCL} when
        fitted to the LQCD and LCSR pseudo data.
    }
    \label{tab:BCL-bfp-LCSR+lQCD}
\end{table}

We emphasize that also the fits to LCSR data only shown in this section are carried out with the
modified BCL parametrization as given in \refeq{modified-BCL} and therefore differ slightly to those obtained in \refsec{lcsr}.\\

We can see that the precision of the extrapolation of the form factors significantly improves by combining
the LCSR and LQCD inputs, especially in the large $q^2$ region, as expected. We are now in a position to revisit
\refeq{fp-residue-analytic} to extract the strong coupling constant from the combined fit. We obtain:
\begin{equation}
    \label{eq:gBstBpi-LCSR+LQCD}
    g_{B^\ast B\pi} = 39.8\pm 1.1\,, 
\end{equation}
where the uncertainties are of parametric origin only. Systematic uncertainties due to higher orders
in the $z$ expansion are not taken into account, and could be sizable due to the magnitude of $z(q^2 = m_{B^*}^2)$.
Our result \refeq{gBstBpi-LCSR+LQCD} agrees well with the lattice determination $g_{B^\ast B\pi} = 45.3 \pm {6.0}$
by the RBC/UKQCD collaboration~\cite{Flynn:2015xna}, but it shows a tension with respect to the recent direct LCSR determination $g_{B^\ast B\pi} = 30.1^{+2.6}_{-2.4}$~\cite{Khodjamirian:2020mlb} at the level of $3.4\,\sigma$.\footnote{
Note that it was already observed in \cite{Khodjamirian:2020mlb} that a significantly larger result arises
when using \refeq{fp-residue-analytic} than when calculating this quantity directly.}
\\

However, we observe that our extrapolation becomes unstable for $q^2 \geq t_-$, \emph{i.e.}, outside the semileptonic phase space:
the $68\%$ probability region for the $f_0$ form factor starts to cover both positive and negative values,
and the central value turns negative just below $q^2 = t_+$.
This finding of negative form factors is inconsistent with a dispersive representation of the form factor.
We suspect the behaviour
to be an artifact of the fit model. Hence, we see no meaningful way to compare our results to the expectation
from the Callan-Treiman type relation in \refeq{Callan-Treiman-type-rel}.\\

\begin{figure}[t]
    \centering
    \includegraphics[width=0.75\textwidth]{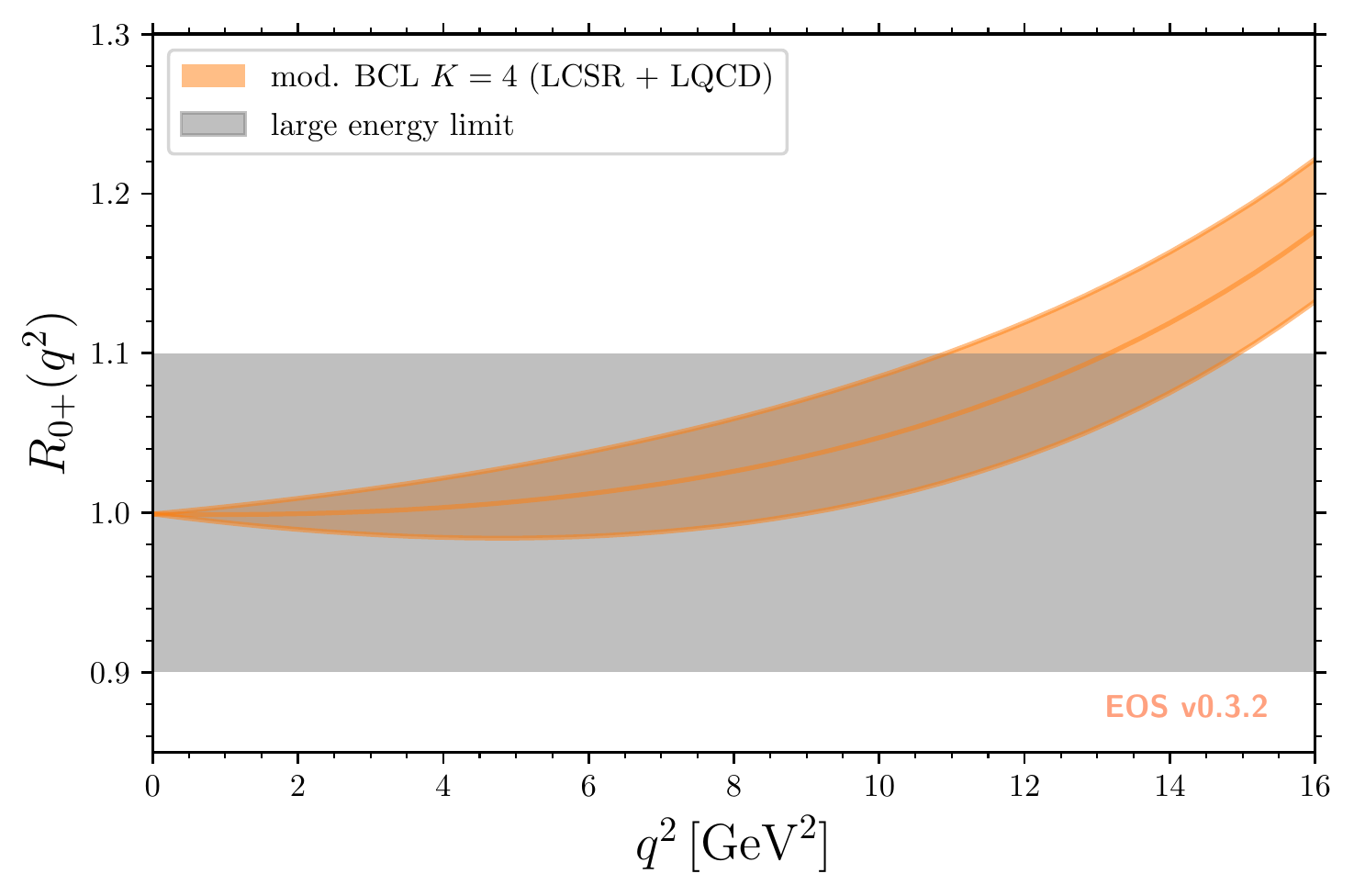}
    \caption{
        The form factor ratio $R_{0+}(q^2)$ interpolated between LCSR and LQCD data and
        compared to the large-energy symmetry limit in the low $q^2$ region.
        }
    \label{fig:le-ratio-LCSR+LQCD}
\end{figure}

In \reffig{le-ratio-LCSR+LQCD} we provide a plot of $R_{0+}(q^2)$ for the form factors interpolating the LCSR and LQCD data.
We find our results to be consistent with the large-energy limit from \refeq{HQ_lim_f0}. Compared to the LCSR-only result, the
range in which the large-energy symmetry limit holds has expanded up to $\simeq 15\,\mathrm{GeV}^2$.\\

A comparison of our results and those in the literature is compiled in \reftab{ff-comparison}.

\begin{table}[b]
    \centering
    \renewcommand{\arraystretch}{1.2}
    \begin{tabular}{ c c c }
        \toprule
        Source           & $f_+(0) = f_0(0)$            & $f_T(0)$         \\
        \midrule
        \multicolumn{3}{c}{Lattice QCD}\\
        \hline 
        Fermilab/MILC \cite{Lattice:2015tia,Bailey:2015nbd}
                         & $0.2\pm 0.2$                     & $0.2\pm 0.2$     \\
        RBC/UKQCD \cite{Flynn:2015mha}
                         & $0.24\pm 0.08$                   & ---  \\
        combination w/ Pade approx. \cite{Gonzalez-Solis:2018ooo}
                         &  $0.265 \pm 0.010 \pm 0.002$ & --- \\
        \midrule
        \multicolumn{3}{c}{Light-cone sum rules}\\
        \midrule
        \text{Duplancic et al.} \cite{Duplancic:2008ix} 
                         & $0.26^{+0.04}_{-0.03}$       & $0.255 \pm 0.035$\\
        \text{Imsong et al.} \cite{Imsong:2014oqa}
                         & $0.31 \pm 0.02$              &  --- \\
        \text{Bharucha} \cite{Bharucha:2012wy}
                         & $0.261^{+0.020}_{-0.023}$              &  --- \\
        \text{Khodjamirian/Rusov} \cite{Khodjamirian:2017fxg}
                         & $0.301 \pm 0.023$            & $0.273 \pm 0.021$  \\
        \text{Gubernari et al.} ($B$ LCDA) \cite{Gubernari:2018wyi}
                         & $0.21 \pm 0.07$              & $0.19 \pm 0.06$\\
        this work        & $0.283\pm 0.027$             & $0.282\pm 0.026$\\
        \midrule
        \multicolumn{3}{c}{Light-cone sum rules + Lattice QCD combination}\\
        \midrule
        this work        &$0.235\pm 0.019$             &$0.235\pm 0.017$\\
        \bottomrule
    \end{tabular}
    \caption{
        Comparison of our results for the form factor normalizations with other QCD-based results in the literature.
        The result of ref.~\cite{Gubernari:2018wyi} is included for completeness, although
        the authors caution that the threshold setting procedure employed in that work
        fails for the $\bar{B}\to \pi$ form factors.
    }
    \label{tab:ff-comparison}
\end{table}

\section{Determination of $|V_{ub}|$ and further phenomenological applications}
\label{sec:Vub-determination}

\subsection{Exclusive $|V_{ub}|$ determination from $\bar{B} \to \pi$ semileptonic decays}

Following the determination of the form factors from LCSRs and lattice QCD input,
we are now in position to extract the magnitude of the CKM matrix element $|V_{ub}|$
from measurements of the $\bar{B} \to \pi \ell^- \bar\nu_\ell$ branching ratio.\\

To this end, we use the world average of the branching ratio as provided by the
HFLAV collaboration~\cite{Amhis:2019ckw}. This average is based on individual
measurements by the BaBar~\cite{delAmoSanchez:2010af,Lees:2012vv} and
Belle~\cite{Ha:2010rf,Sibidanov:2013rkk} collaborations. The world
average is provided in terms of 13 bins of the squared momentum transfer $q^2$,
with identical bin sizes. Within the averaging process, HFLAV accounts for
shared systematic correlations among the individual measurements.\\

\begin{figure}[t]
    \centering
    \includegraphics[width=0.9\textwidth]{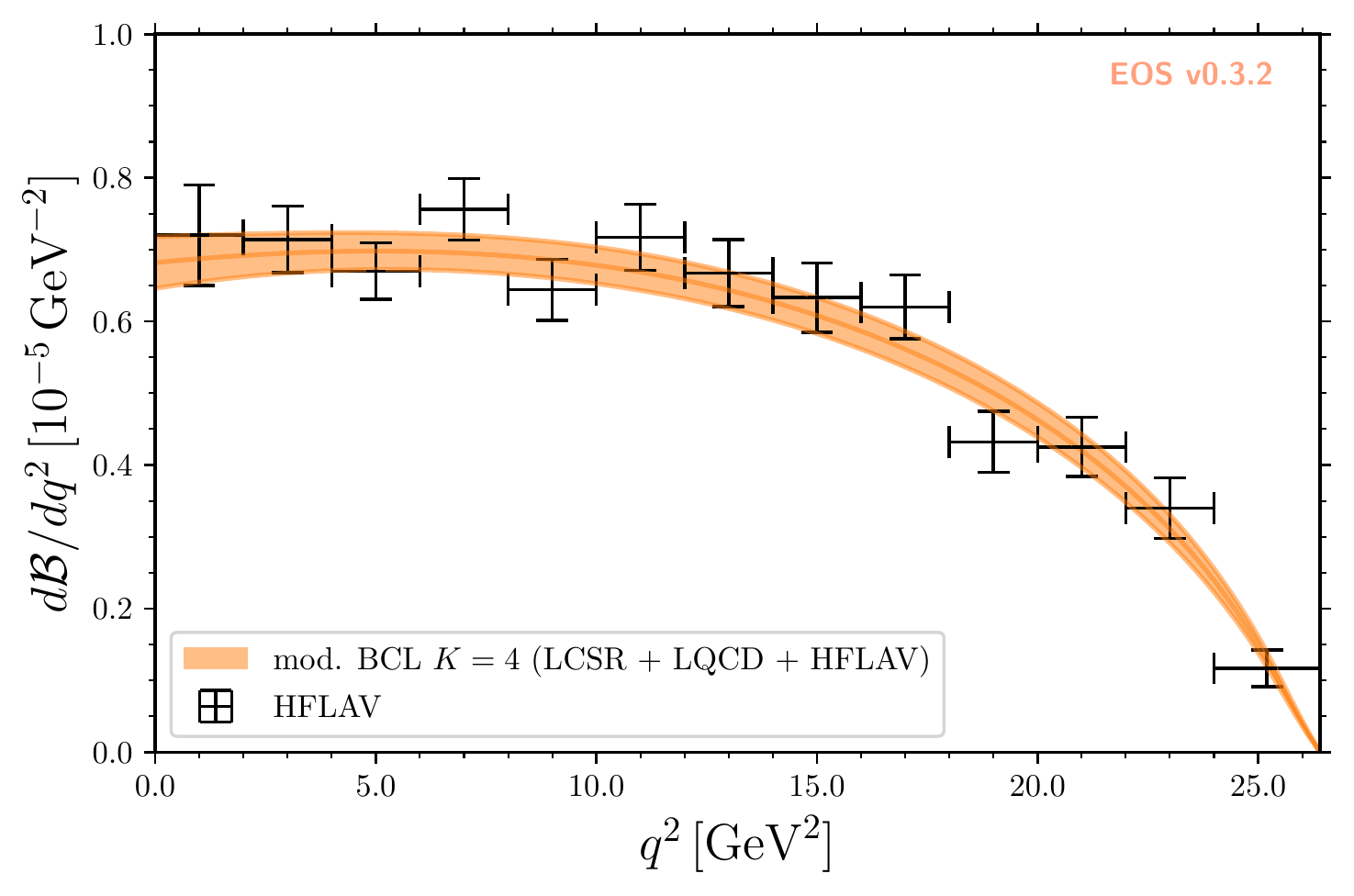}
    \caption{Differential branching ratio for $\bar{B}^0 \to \pi^+ \ell^- \bar{\nu}_\ell$ decay as obtained from the combined fit to LCSR and lattice QCD inputs and experimental data, compared to the HFLAV average of the experimental data.
    }
    \label{fig:diff-br}
\end{figure}

A visual representation of this data, which we provide in \reffig{diff-br},
shows that the highest relative experimental precision is achieved in
for intermediate $q^2$, \emph{i.e.}, in a region that is not reliably accessible
with LCSRs and not yet accessible with lattice QCD simulations. Consequently,
our efforts to obtain high-precision determinations of the form factors at
intermediate $q^2$ through interpolation of the respective theory results
is of high importance to the $|V_{ub}|$ determination. This is nicely illustrated
in Fig.~3 of ref.~\cite{Becher:2005bg}.\\

\begin{table}[t]
    \def\arraystretch{1.3}
    \centering
    \begin{tabular}{ c r @{$^{+}_{-}$} l r @{$^{+}_{-}$} l r @{$^{+}_{-}$} l}
        \toprule
        \multirow{2}{*}{\diagbox[height=2\line]{param.}{method}} & \multicolumn{4}{c}{LCSR+LQCD} & \multicolumn{2}{c}{LCSR only}  \\
                            & \multicolumn{2}{c}{$K=3$} & \multicolumn{2}{c}{$K=4$}     & \multicolumn{2}{c}{$K=3$} \\
        \midrule
        $10^{-3}\times |V_{ub}|$&  $ 3.80 $   & $ ^{0.14}_{0.14}$    &  $ 3.77 $  &   $ ^{0.15}_{0.15} $  & $  3.28 $  & $ ^{0.33}_{0.28}$ \\
        $f_+(0)$                &  $  0.248 $ & $ ^{0.009}_{0.009} $ & $  0.246 $ & $ ^{0.009}_{0.009} $ & $  0.284 $ & $^{0.025}_{0.025}$ \\
        $b^+_1$                 &  $- 2.13 $  & $ ^{0.19}_{0.19} $   & $- 2.10 $  & $ ^{0.22}_{0.21} $   & $- 1.91 $  & $^{0.31}_{0.30}$ \\
        $b^+_2$                 &  $- 0.82 $  & $ ^{0.54}_{0.55} $   & $  0.23 $ & $ ^{0.87}_{0.87} $    & $- 1.42 $  & $^{0.85}_{0.89}$ \\
        $b^+_3$                 &\multicolumn{2}{c}{---}             & $- 3.0$   & $ ^{2.8}_{2.8} $ &\multicolumn{2}{c}{---} \\
        \midrule
        $\chi^2/\mathrm{d.o.f}$ & \multicolumn{2}{c}{$\sim 32.33/34$} & \multicolumn{2}{c}{$\sim 29.30/31$} & \multicolumn{2}{c}{$\sim 10.72/17$} \\
        $p$ value               & \multicolumn{2}{c}{$\sim 55\%$}     & \multicolumn{2}{c}{$\sim 55\%$}     & \multicolumn{2}{c}{$\sim 87\%$} \\
        \bottomrule
    \end{tabular}
    \caption{
        Results from the three fits to combinations of fit models and data sets as described in the text.
        We provide the median values and central intervals at $68\%$ probability for the one-dimensional
        marginalized posterior distributions.
    }
    \label{tab:goodness-of-fit-Vub}
\end{table}

Our analysis is set up in the same way as in \refsec{interpolation}.
We stress that this means that we exclusively fit using the modified
BCL parametrization. As the only modification with respect to \refsec{interpolation}
we include the HFLAV average as part of the likelihood. 
The theory prediction for the $\bar{B}\to \pi\ell^-\bar\nu_\ell$ branching ratio does not depend on the form factor
$f_T$ in the SM, which we assume for our fit. For $\ell=e,\mu$ the branching ratio is only very weakly
dependent on the form factor $f_0$, which contributes measurably only for $q^2 \lesssim 1\,\GeV^2$.
Additionally, the predictions for $f_0$ are affected by interaction between the experimental constraint
on $f_+$ and the theoretical correlations between $f_+$ and $f_0$.
As a consequence, we present our results as one-dimensional marginalized posterior
distributions only for $|V_{ub}|$ and the parameters describing the $f_+$ form factor.
We carry out fits to the LCSR pseudo data only in the $K=3$ fit model as well as
combined fits to the LCSR + lattice QCD inputs in the $K=3$ and $K=4$ models.
In all cases we find a good fit, with $p$ values in excess of $55\%$. While
the $K=4$ fit model to LCSR + lattice QCD inputs does not provide a significantly improved goodness of fit,
we still adopt it as our nominal fit model. Our reasoning is that this model
can account for additional systematic uncertainties inherent to the extrapolation process,
which slightly increases the uncertainty of the $|V_{ub}|$ extraction.
The smallness of the difference in the $K=3$ and $K=4$ uncertainties seems to indicate
that systematic uncertainties are under reasonable control.
Summaries of the one-dimension marginalized posteriors in terms of their
median values and central $68\%$ probability intervals are provided in \reftab{goodness-of-fit-Vub}.
\\

\begin{table}[t]
    \centering
    \renewcommand{\arraystretch}{1.2}
    \begin{tabular}{ c c }
        \toprule
        Source           & $10^{-3}\times |V_{ub}|$ \\
        \midrule
        \multicolumn{2}{c}{LQCD}\\
        \hline 
        Fermilab/MILC \cite{Lattice:2015tia,Bailey:2015nbd}
                         & $3.72 \pm 0.16$ \\
        RBC/UKQCD \cite{Flynn:2015mha}
                         & $3.61 \pm 0.32$ \\
        combination w/ Pade approx. \cite{Gonzalez-Solis:2018ooo}
                         & $3.53 \pm 0.08_\text{stat} \pm 0.06_\text{syst}$ \\
        HFLAV \cite{Amhis:2019ckw}
                         & $3.70 \pm 0.10_\text{stat} \pm 0.12_\text{syst}$ \\
        \midrule
        \multicolumn{2}{c}{LCSR}\\
        \midrule
        \text{Duplancic et al.} \cite{Duplancic:2008ix} 
                         & $3.5 \pm 0.4\pm 0.2 \pm 0.1$ \\
        \text{Imsong et al.} \cite{Imsong:2014oqa}
                         & $3.32^{+0.26}_{-0.22}$ \\
        this work        & $3.28^{+0.33}_{-0.28}$ \\
        \midrule
        \multicolumn{2}{c}{LCSR + LQCD}\\
        \midrule
        HFLAV \cite{Amhis:2019ckw}
                         & $3.67 \pm 0.09_\text{stat} \pm 0.12_\text{syst}$ \\
        this work        & $3.77 \pm 0.15 $ \\
        \bottomrule
    \end{tabular}
    \caption{Comparison of the $|V_{ub}|$ CKM matrix element determinations from the $\bar{B}\to \pi \ell^- \bar{\nu}_\ell$ decays, using  QCD-based form factor predictions.}
    \label{tab:Vub-comparison}
\end{table}

We find that the LCSR-only fit yields a $|V_{ub}|$ result that is slightly smaller than
the LCSR + lattice LQCD results by approximately more than one sigma.
The latter results for $K=3$ and $K=4$ are in mutual agreement.
This is not surprising, given the shift in $f_+(0)$ between these two scenarios, which
is already discussed in \refsec{interpolation}.
The results for $K=3$ and $K=4$ for fit to LCSR + lattice LQCD results are perfectly compatible with each other.
Our nominal result is obtained from the fit to LCSR + lattice QCD input with $K=4$, and
reads
\begin{equation}
    |V_{ub}|^{\bar{B}\to \pi}_\text{LCSR+LQCD} = (3.77 \pm 0.15) \cdot 10^{-3}.
\end{equation}

The apparent slight tension between $f_+(0)$ obtained from the fit to LCSR data only and the fit to LCSR+LQCD data, as previously discussed in~\refsec{interpolation}, persists here as well. It translates to a reasonable agreement between
the determinations of $|V_{ub}|$ at the $1.4\,\sigma$ level. We find a very good fit using the combined LCSR and LQCD data, with $\chi^2/\mathrm{d.o.f.}\sim 1$, and a $p$ value of $\sim 55\%$ at the best-fit point. Adding information on the
form factor shape through the HFLAV average of the experimental data does not affect our results for the BCL parameters compared to results of the theory-only fit in \refsec{interpolation}.
This result exhibits a slight tension with the BLNP and GGOU determinations, in both cases at the
$\approx 2\,\sigma$ level. However, it is in very good agreement with the
recent method-averaged result by the Belle collaboration as given in \refeq{Vub-Belle-inclusive-2020}.
Here the tension reduces to $\approx 1\,\sigma$ only.\\

We compare our results for $|V_{ub}|$ with other methods in \reftab{Vub-comparison} and  
in \reffig{norm-br} give our Standard Model prediction for the differential decay rate of $\bar{B} \to \pi \ell^- \nu_\ell$  divided by $|V_{ub}|$ for the electron and tau lepton final states.\\

The normalized branching ratios obtained with the use of the theory-only form factors from \refsec{interpolation} yield
\begin{equation}
    \begin{split}
        \mathcal{B}(\bar{B} \to \pi \mu^- \bar{\nu}_\mu)  & = \left(9.6^{+1.0}_{-1.0}\right)\times |V_{ub}|^2\,,\\
        \mathcal{B}(\bar{B} \to \pi \tau^- \bar{\nu}_\tau) & = \left(6.7^{+0.6}_{-0.5}\right)\times |V_{ub}|^2\,.
    \end{split}
\end{equation}

\subsection{Lepton-Flavour Universality Ratio}

Next, we make predictions for Standard Model observables for the $\bar{B} \to \pi \ell^- \bar\nu_\ell$ decay results as obtained in \refsec{interpolation}.\\

\begin{figure}[t]
    \centering
    \includegraphics[width=0.9\textwidth]{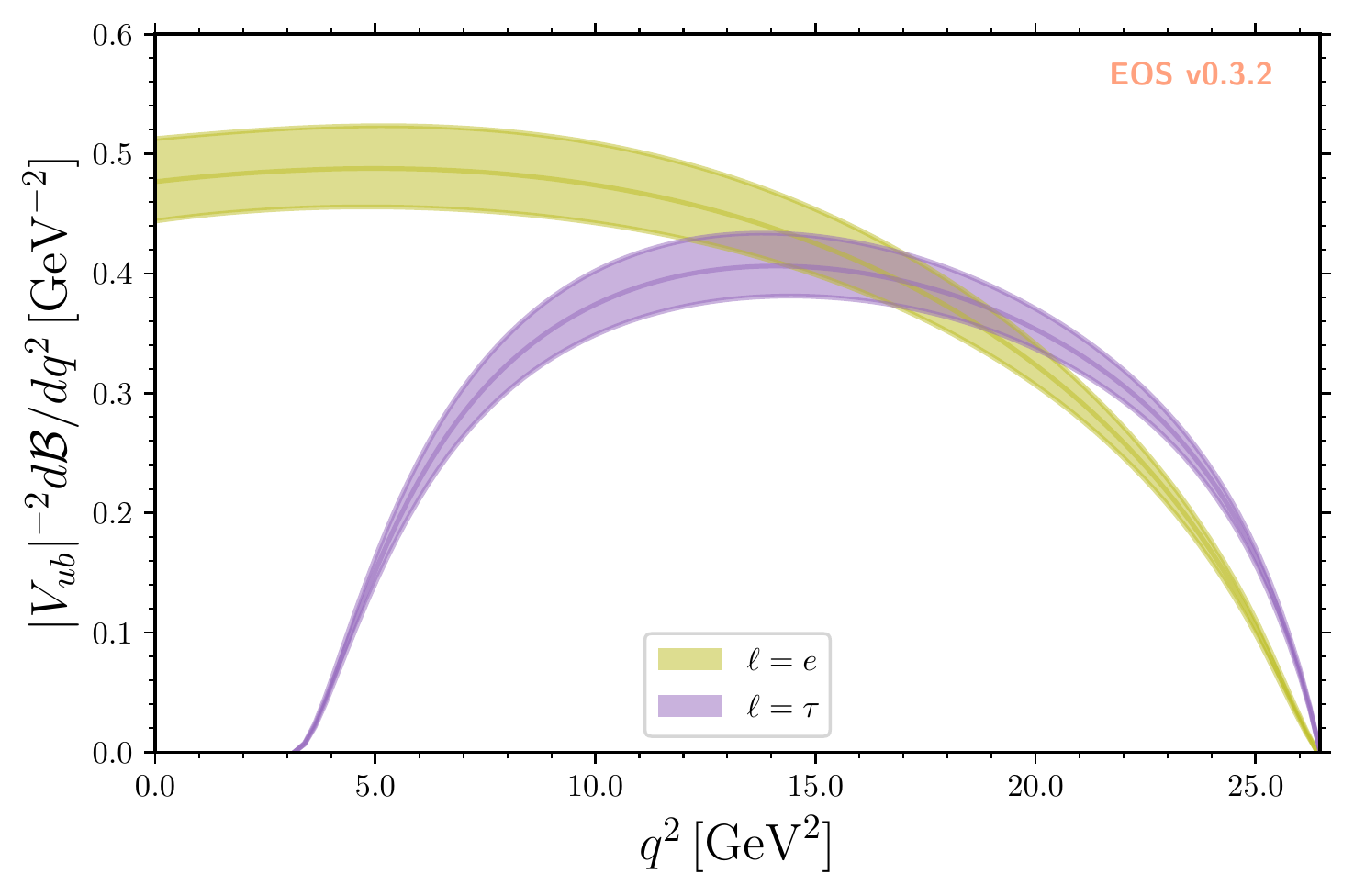}
    \caption{Our Standard Model prediction of the differential decay rate of $\bar{B} \to \pi \ell^- \bar{\nu}_\ell$  divided by $|V_{ub}|$ for the electron and tau final states. The form factors are obtained from a fit to both LCSR and
    LQCD inputs using modified BCL parametrization in the $K=4$ fit model.
    }
    \label{fig:norm-br}
\end{figure}

In light of hints for LFU violating effects in $\bar{B}\to D^{(*)}\ell^-\bar{\nu}_\ell$ decays~\cite{Amhis:2019ckw,Bernlochner:2021vlv},
we investigate the LFU-probing observable for $\bar{B}\to\pi\ell^-\bar\nu_\ell$ decays:
\begin{equation}
    \label{eq:R-pi}
    \begin{aligned}
        R_{\pi}
            & = \frac{\Gamma (\bar{B} \to \pi \tau^- \bar{\nu}_{\tau})}{\Gamma (\bar{B} \to \pi \ell^- \bar{\nu}_{\ell})}
              = \frac{\int_{m_{\tau}^2}^ {q_{\rm max}^2} d\Gamma (\bar{B} \to \pi \tau^- \bar{\nu}_{\tau})/dq^2}{\int_{m_\ell^2}^ {q_{\rm max}^2} d\Gamma (\bar{B} \to \pi \ell^- \bar{\nu}_{\ell})/dq^2}, \qquad ( \ell = e,\mu)\,.
    \end{aligned}
\end{equation}
In the Standard model, predictions for $R_\pi$ involve only two out of three form factors, $f_+$ and $f_0$.
Using our results from the form factor fit with $K=4$ as obtained in \refsec{interpolation} we find:
\begin{eqnarray}
    R_{\pi}\big|_{\text{LCSR+LQCD}} = 0.699^{+0.022}_{-0.020}\,.
\end{eqnarray}
The central values for $R_\pi$ as predicted from the $K=3$ and $K=5$ fits fall entirely within the above uncertainties.
We also show the differential branching ratios for the tauonic and light-lepton modes individually in
\reffig{norm-br}.\\

It is important to stress that for a precise determination of $R_\pi$ knowledge of the scalar form factor $f_0(q^2)$
is key. To demonstrate this, we disentangle the contributions to the tauonic decay width stemming from each of the form factors:
\begin{equation}
    R_\pi \equiv R_{\pi}^+ + R_{\pi}^0\,,
\end{equation}
corresponding to the $|f_+|^2$ and $|f_0|^2$ contributions, respectively. We find
\begin{equation}
    \begin{aligned}
    R_{\pi}^{+}\big|_{\text{LCSR+LQCD}}
        & = 0.476^{+0.014}_{-0.013}\,, &
    R_{\pi}^{0}\big|_{\mathrm{LCSR+LQCD}}
        & = 0.224^{+0.014}_{-0.013}\,.
    \end{aligned}
\end{equation}
Although the $f_0$ contribution is half the size of the $f_+$ contribution, its relative uncertainty
is about two times as large as the one of the $f_+$ term. This illustrates the importance of
accurately predicting both of the form factors for this LFU probe.\\

In \reftab{R-pi-comparison} we provide a comparison of our results with the available determinations of $R_\pi$ in
the literature. We find that our prediction is 
in very good agreement at or below the $1\,\sigma$ level with the previous determinations provided in refs.~\cite{Flynn:2015mha,Becirevic:2020rzi}.
A minor exception is the prediction of ref.~\cite{Bernlochner:2015mya}, which is in agreement  with our result at the
$2\,\sigma$ level.\\

Although the LFU ratio $R_\pi$ is $|V_{ub}|$-independent, it could, on the other hand, be sensitive to potential new physics effects in $\bar{B} \to \pi \tau^- \bar{\nu}_{\tau}$ decay due to the presence of new scalar currents and/or electroweak symmetry breaking effects associated with the large mass of the $\tau$ lepton.
Hence, $R_\pi$ is a very interesting candidate for future measurements.
To date, there is a single experimental result by the Belle collaboration~\cite{Hamer:2015jsa}.
It is obtained from an upper limit on the branching ratio of $\bar{B}\to \pi\tau^-\bar\nu_\tau$,
which has not yet been observed. This result reads:
\begin{equation}
    R_\pi\big|_{\text{Belle}} = 1.05\pm 0.51\,,
\end{equation}
which is in agreement with our prediction.\\

\begin{table}[t]
    \centering
    \begin{tabular}{c | c r @{$\pm$} l r @{$\pm$} l r @{$\pm$} l}
        \toprule
        \multirow{2}{*}{Th. only}
                & source
                & \multicolumn{2}{c}{RBC/UKQCD~\cite{Flynn:2015mha}}
                & \multicolumn{2}{c}{Bečirević et al.~\cite{Becirevic:2020rzi}}
                & \multicolumn{2}{c}{this work}\\
                & $R_\pi$
                & 0.69  & 0.19
                & 0.78  & 0.10
                & 0.699 & 0.022 \\
        \midrule
        \multirow{2}{*}{Th. + Exp.}
                & source
                & \multicolumn{2}{c}{Bernlochner~\cite{Bernlochner:2015mya}}
                & \multicolumn{2}{c}{Bečirević et al.~\cite{Becirevic:2020rzi}}
                & \multicolumn{2}{c}{this work}\\
                & $R_\pi$
                & 0.641 & 0.016
                & 0.66  & 0.02
                & 0.688 & 0.014 \\
        \bottomrule
    \end{tabular}
    \caption{Comparison of theory results for the LFU ratio $R_\pi$ in the literature.}
    \label{tab:R-pi-comparison}
\end{table}

\subsection{Angular observables and polarizations in $\bar{B}\to \pi\ell^-\bar\nu_\ell$}

We can now use our results for the form factors to predict the two angular observables in the
two-fold differential decay rate of $\bar{B}\to \pi\ell^-\bar\nu_\ell$ as well as the lepton polarization
in these decays.\\

We begin with the discussion of the forward-backward asymmetry in the Standard model. The integrated normalized forward-backward asymmetry is defined as
\begin{equation}
    A_{\rm FB}^\ell
        =   \frac{1}{\Gamma(\bar{B} \to \pi \ell^- \bar{\nu}_\ell)}
            \int_{m_\ell^2}^{q_{\rm max}^2} dq^2\left [\int_{-1}^0 -  \int_{0}^{-1}  \right ] d\cos\theta_\ell \frac{d \Gamma^2(\bar{B} \to \pi \ell^- \bar{\nu}_\ell) }{ dq^2 d\cos\theta_\ell}.
\end{equation}
The forward-backward asymmetry arises from interference of the timelike polarization with the longitudinal polarization
of the dilepton final state. The asymmetry is proportional to the mass of the charged lepton.
Hence, $A_\text{FB}$ is small for the $\ell=e,\mu$, which makes it very sensitive to BSM effects
that could lift the helicity suppression.
With our results for the form factors from \refsec{interpolation} we obtain in the SM
\begin{equation}
    \begin{split}
        A_{\mathrm{FB}}^{\mu}    & = -0.0048\pm 0.0003\,,\\
        A_{\mathrm{FB}}^\tau     & = -0.259 \pm 0.004\,.
    \end{split}
\end{equation}
We do not provide the SM prediction for the electron mode, since it is indistinguishable from zero.
Our results are in reasonable agreement with the RBC/UKQCD results~\cite{Flynn:2015mha}, but are
more precise.\footnote{Note that the convention for lepton helicity angle in ref.~\cite{Flynn:2015mha}
differs from our by a sign, which is also affecting the overall sign of $A_\text{FB}$.}\\

The flat term $F_H$~\cite{Bobeth:2007dw,Bouchard:2013mia} is another
observables that arises in the normalized angular distribution.
In the SM it is proportional to the lepton mass and therefore small.
This makes it an appropriate candidate for a BSM probe, too. It can be defined as
\begin{equation}
    F_H^\ell = 1 + \frac{2}{3} \frac{1}{\Gamma(\bar{B} \to \pi \ell^- \bar\nu_\ell)} \frac{d^2}{d (\cos\theta)^2} \left[\frac{d\Gamma(\bar{B} \to \pi \ell^- \bar\nu_\ell)}{d\cos\theta}\right] = 1 + \frac{2}{3} C_F^\ell\,,
\end{equation}
and is therefore related to the convexity parameter $C_F^\ell$.
With our results for the form factors from \refsec{interpolation} we obtain in the SM
\begin{equation}
    F_H^{\mu} =0.0024 \pm 0.0001; \qquad F_H^{\tau} = 0.134\pm 0.003\,.
\end{equation}
We do not provide the SM prediction for the electron mode, since it is indistinguishable from zero.

As a final BSM probe we investigate the integrated normalized $\tau$ polarization asymmetry,
which can be expressed as
\begin{equation}
    P^{\tau} = \frac{\Gamma(\bar{B} \to \pi \tau^-_\uparrow \bar\nu_\tau) - \Gamma(\bar{B} \to \pi \tau^-_\downarrow \bar{\nu}_\tau)}{\Gamma( \bar{B} \to \pi \tau^- \bar{\nu}_\tau)}\,,
\end{equation}
where $\tau_{\uparrow,\downarrow}$ denotes the tau helicities $\lambda_\tau = \pm 1/2$. 
With our results for the form factors from \refsec{interpolation} we obtain
\begin{equation}
    P^{\tau} = -0.21 \pm 0.02.
\end{equation}

\section{Conclusions}

We study the $\bar{B}\to \pi$ form factors and use their numerical results from QCD-based methods
to update the exclusive determination of $|V_{ub}|$ from $\bar{B} \to \pi$ semileptonic decays
and the SM predictions of a number of phenomenologically interesting observables.\\

We begin by revisiting the determination of the form factors using light-cone sum rules with $\pi$ distribution amplitudes.
Our analysis includes the full set of local $\bar{B}\to \pi$ form factors of dimension-three currents.
For the first time, we apply a threshold setting procedure based on Bayesian inference to the full set of
these form factors. Beside the thresholds, we are also able to infer a value for the (unphysical) Borel parameter
that is mutually compatible among all three form factors.\\

Our results for the form factors, obtained at small momentum transfer $q^2$, are then extrapolated to large $q^2$ by applying a standard BCL parametrization. We show that this extrapolation agrees well with precise lattice QCD results
for the form factor $f_+$ and $f_T$. However, in order to ensure
good agreement also for the form factor $f_0$, we find that its parametrization needs to be modified.
We stress that for precise and consistent predictions the correct treatment of the correlations in the pseudo data
points is crucial, especially between $f_+$ and $f_0$; this is sometimes overlooked in the literature.
We provide correlated results for the normalization and shape parameters of all form factors, including their correlations
through ancillary machine-readable data files.\\

Our predictions for the form factors agree very well with measurements of the $q^2$ spectrum of the
semileptonic decay $\bar{B}^0\to \pi^+\ell^-\bar\nu_\ell$. Using its current world average we determine
$|V_{ub}| = (3.77 \pm 0.15)\cdot 10^{-3}$. Our result is in good agreement with the most recent
inclusive determination by Belle at the $1\,\sigma$ level, which removes the long-standing tension
between inclusive and exclusive $|V_{ub}|$
determinations.\\

The form factors at hand, we also compute SM predictions for a number of phenomenologically interesting observables,
such as the lepton-flavour universality ratio $R_\pi$, the leptonic forward-backward asymmetry $A_\text{FB}$, the flat term $F_H$ and the $\tau$ polarization $P_\tau$. 
Based on our precise and correlated results for the form factors we obtain
very precise predictions of the aforementioned observables. Their relative
uncertainties range from $\approx 10\%$ for the branching ratios to
about $4\%$ for some of the normalized observables.
We are looking forward to precision measurements of these observables by the Belle experiment,
which will further constrain the form factors and probe the SM at a precision level.

\acknowledgments
We thank Florian Bernlochner for insightful discussions and Kenji Nishiwaki for participating in the early stages of this work.
D.v.D.~is supported by the DFG within the Emmy Noether Programme under grant DY-130/1-1
and the DFG Collaborative Research Center 110 ``Symmetries and the Emergence of Structure in QCD''. B.M.~and D.L.~have
been supported by the European Union through the European Regional Development Fund – the Competitiveness and Cohesion
Operational Programme (KK.01.1.1.06) and by the Croatian Science Foundation (HRZZ) project ``Heavy hadron decays and
lifetimes'' IP-2019-04-7094.  Sponsorship has been also provided by the Alexander von Humboldt Foundation in the
framework of the Research Group Linkage Programme, funded by the German Federal Ministry of
Education and Research. B.M.~would like to express her gratitude to Wolfram Research for providing a free licence to use {\fontfamily{cmtl}\selectfont Mathematica} at home during the COVID-19 pandemic lockdown.
B.M.~and D.v.D.~would like to express special thanks to the Mainz Institute for Theoretical Physics (MITP)
of the Cluster of Excellence PRISMA+ (Project ID 39083149) for its hospitality and support.

\appendix

\section{Supplementary information}
\label{app:lcsr-corr}

In \reftab{corr_LCSR} we provide the correlation matrix of the form factor pseudo data set obtained from LCSRs.

\begin{table}[H]
    \tiny
    \centering
    \setlength{\tabcolsep}{3pt}
    \begin{tabular}{ c | c c c c c c c c c c c c c c }
        $\mathcal{F}$ & $f_+(-10)$ & $f_+(-5)$ & $f_+(0)$ & $f_+(5)$ & $f_+(10)$ & $f_0(-10)$ & $f_0(-5)$ & $f_0(5)$ & $f_0(10)$ & $f_T(-10)$ & $f_T(-5)$ & $f_T(0)$ & $f_T(5)$ & $f_T(10)$ \\
        \hline
        $f_{+}(-10)$ & 1.00 &    0.70 &    0.46 &    0.32 &    0.28 &    0.78 &    0.71 &    0.33 &    0.31 &    0.33 &    0.43 &    0.41 &    0.39 &    0.40\\
        $f_{+}(-5)$ & 0.70 &    1.00 &    0.68 &    0.58 &    0.55 &    0.68 &    0.81 &    0.58 &    0.56 &    0.45 &    0.62 &    0.59 &    0.56 &    0.57\\
        $f_{+}(0)$ & 0.46 &    0.68 &    1.00 &    0.63 &    0.62 &    0.42 &    0.67 &    0.63 &    0.60 &    0.43 &    0.60 &    0.58 &    0.55 &    0.56\\
        $f_{+}(5)$ & 0.32 &    0.58 &    0.63 &    1.00 &    0.63 &    0.28 &    0.56 &    0.62 &    0.60 &    0.40 &    0.56 &    0.55 &    0.52 &    0.54\\
        $f_{+}(10)$ & 0.28 &    0.55 &    0.62 &    0.63 &    1.00 &    0.24 &    0.53 &    0.62 &    0.62 &    0.40 &    0.56 &    0.55 &    0.54 &    0.56\\
        $f_{0}(-10)$ & 0.78 &    0.68 &    0.42 &    0.28 &    0.24 &    1.00 &    0.69 &    0.29 &    0.28 &    0.30 &    0.40 &    0.38 &    0.36 &    0.37\\
        $f_{0}(-5)$ & 0.71 &    0.81 &    0.67 &    0.56 &    0.53 &    0.69 &    1.00 &    0.57 &    0.54 &    0.44 &    0.61 &    0.58 &    0.55 &    0.56\\
        $f_{0}(5)$ & 0.33 &    0.58 &    0.63 &    0.62 &    0.62 &    0.29 &    0.57 &    1.00 &    0.61 &    0.40 &    0.56 &    0.55 &    0.53 &    0.54\\
        $f_{0}(10)$ & 0.31 &    0.56 &    0.60 &    0.60 &    0.62 &    0.28 &    0.54 &    0.61 &    1.00 &    0.40 &    0.56 &    0.55 &    0.53 &    0.55\\
        $f_{T}(-10)$ & 0.33 &    0.45 &    0.43 &    0.40 &    0.40 &    0.30 &    0.44 &    0.40 &    0.40 &    1.00 &    0.69 &    0.46 &    0.34 &    0.32\\
        $f_{T}(-5)$ & 0.43 &    0.62 &    0.60 &    0.56 &    0.56 &    0.40 &    0.61 &    0.56 &    0.56 &    0.69 &    1.00 &    0.66 &    0.58 &    0.57\\
        $f_{T}(0)$ & 0.41 &    0.59 &    0.58 &    0.55 &    0.55 &    0.38 &    0.58 &    0.55 &    0.55 &    0.46 &    0.66 &    1.00 &    0.62 &    0.62\\
        $f_{T}(5)$ & 0.39 &    0.56 &    0.55 &    0.52 &    0.54 &    0.36 &    0.55 &    0.53 &    0.53 &    0.34 &    0.58 &    0.62 &    1.00 &    0.64\\
        $f_{T}(10)$ & 0.40 &    0.57 &    0.56 &    0.54 &    0.56 &    0.37 &    0.56 &    0.54 &    0.55 &    0.32 &    0.57 &    0.62 &    0.64 &    1.00
    \end{tabular}
    \caption{\normalsize The correlation matrix between the LCSR form factors pseudo data points. The momentum transfer is given as the argument in units of $\mathrm{GeV}^2$.}
    \label{tab:corr_LCSR}
\end{table}

\begin{figure}[h]
    \centering
    \begin{tabular}{cc}
        \includegraphics[width=.49\textwidth]{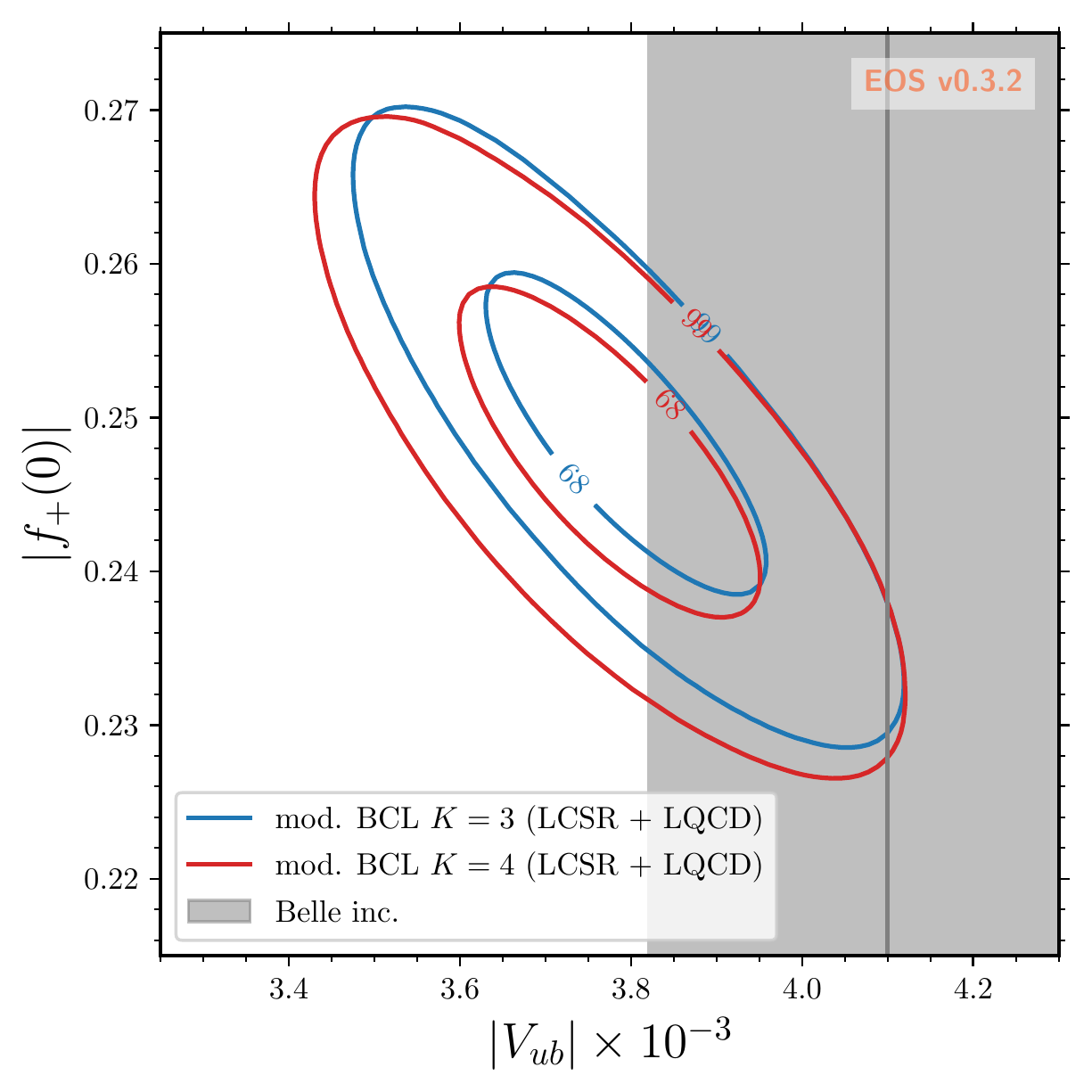} &
        \includegraphics[width=.49\textwidth]{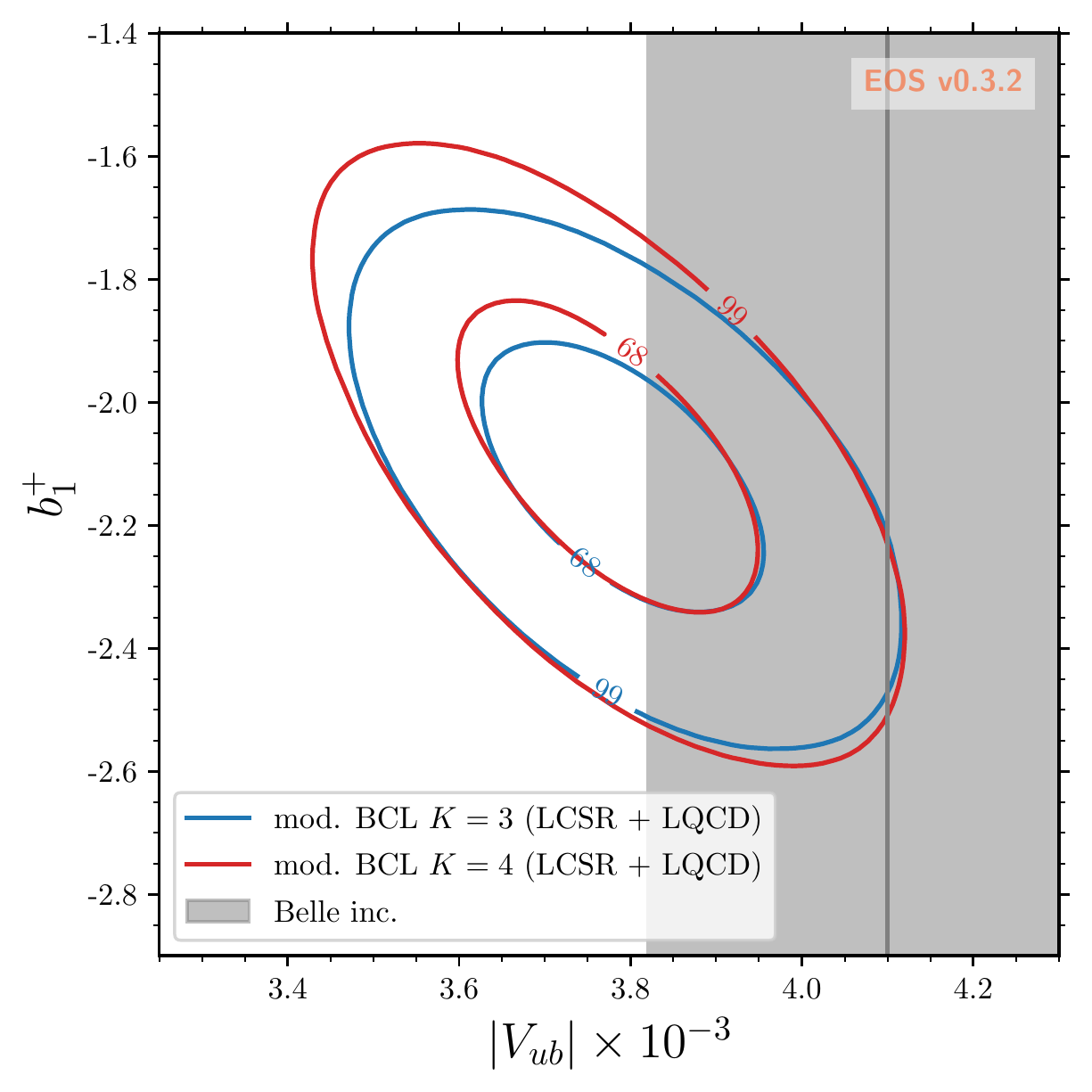} \\
        \includegraphics[width=.49\textwidth]{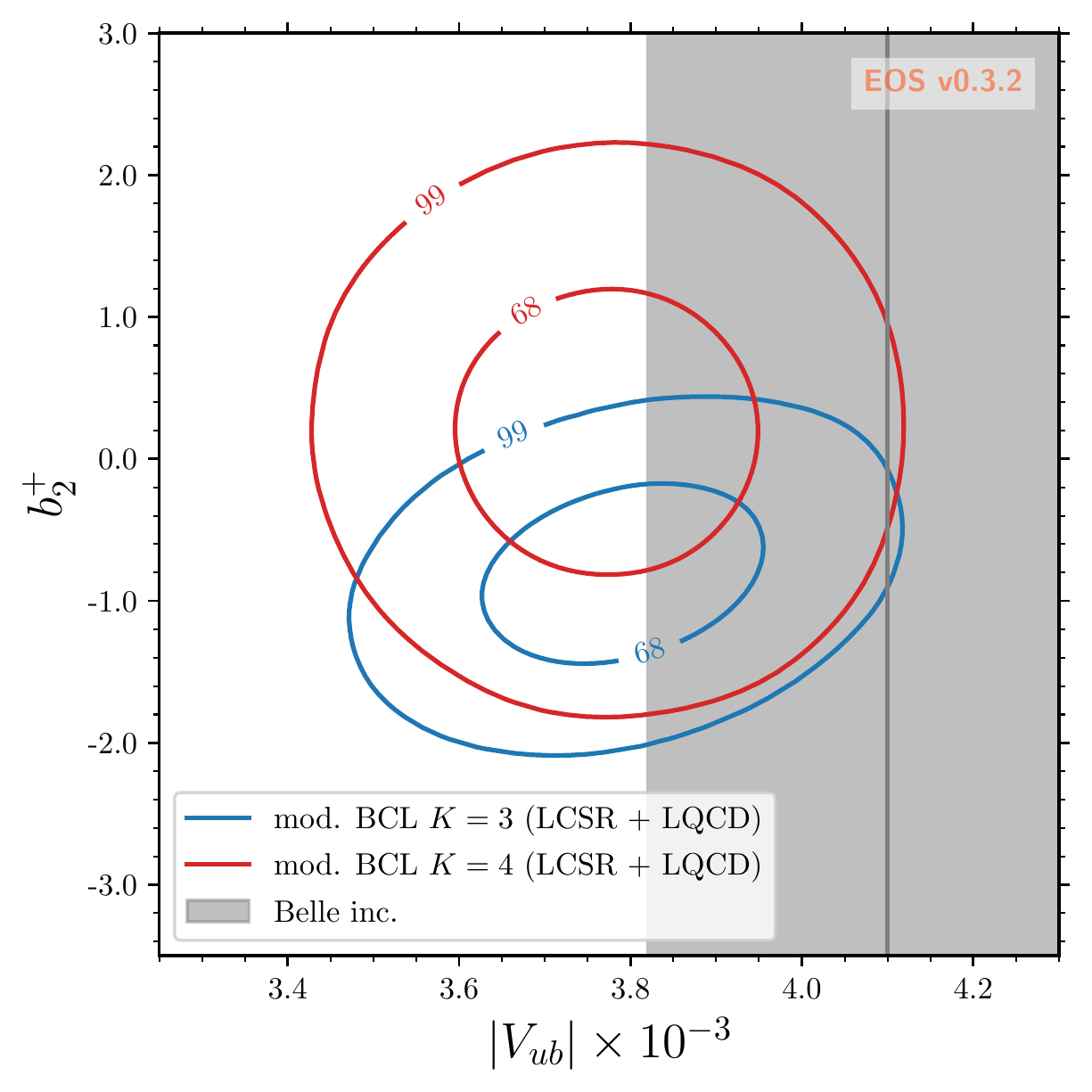} &
        \raisebox{10.\bigskipamount}{
        \begin{minipage}[b]{.45\textwidth}
            \caption{
            Contours of the two-dimensional marginalised posterior distributions of the $K=3$ and $K=4$ fits
            to both theory inputs and experimental data at $68\%$ and $99\%$ probability. For a description
            of the fits, see \refsec{Vub-determination}. The most recent Belle result~\cite{Cao:2021xqf} for the inclusive $|V_{ub}|$ determination is shown as a gray line (central value) and band (uncertainty).
            }
            \label{fig:Vub-parametric}    
        \end{minipage}
        }
    \end{tabular} 
\end{figure}

In \reffig{Vub-parametric} we provide contours of the two-dimensional marginalised posterior distributions that involve $|V_{ub}|$ in the $K=3$ and $K=4$ scenarios of the fits to theory inputs and experimental data.  

\newpage

\bibliographystyle{JHEP}
\bibliography{references}

\end{document}